\documentclass[twoside,twocolumn,english,aps, prx, superscriptaddress, amsfonts, amssymb, amsmath, a4paper, reprint]{revtex4-1}
\usepackage{lmodern}

\usepackage[T1]{fontenc}
\usepackage[utf8]{inputenc}
\usepackage{color}
\usepackage{babel}
\usepackage{amsmath}
\usepackage{amssymb}
\usepackage{graphicx}
\usepackage[all]{xy}
\usepackage[unicode=true,
 bookmarks=false,
 breaklinks=true,pdfborder={0 0 0},pdfborderstyle={},backref=false,colorlinks=true]
 {hyperref}
\hypersetup{pdftitle={Relaxation Dynamics of Spin Qubits in Quantum Dots},
 pdfauthor={Dan Cogan}}


\let\oldxymatrix\xymatrix
\renewcommand{\xymatrix}{\oldxymatrix @C=0.3em @R=0.5em}
\def\arc{\ar@/^2ex/}   


\begin{document}
\title{Optical Phase Measurement Using a Deterministic Source\\
of Entangled Multi-photon States}
\author{G. Peniakov}
\affiliation{The Physics Department and the Solid State Institute, Technion–Israel
Institute of Technology, 3200003 Haifa, Israel}
\author{Z.-E. Su}
\affiliation{The Physics Department and the Solid State Institute, Technion–Israel
Institute of Technology, 3200003 Haifa, Israel}
\author{A. Beck}
\affiliation{The Physics Department and the Solid State Institute, Technion–Israel
Institute of Technology, 3200003 Haifa, Israel}
\author{D. Cogan}
\affiliation{The Physics Department and the Solid State Institute, Technion–Israel
Institute of Technology, 3200003 Haifa, Israel}
\author{O. Amar}
\affiliation{The Physics Department and the Solid State Institute, Technion–Israel
Institute of Technology, 3200003 Haifa, Israel}
\author{D. Gershoni}
\email{dg@physics.technion.ac.il}

\affiliation{The Physics Department and the Solid State Institute, Technion–Israel
Institute of Technology, 3200003 Haifa, Israel}
\begin{abstract}
Precision measurements of optical phases have many applications in
science and technology. Entangled multi-photon states have been suggested
for performing such measurements with precision that significantly
surpasses the shot-noise limit. Until recently, such states have been
generated mainly using spontaneous parametric down-conversion – a
process which is intrinsically probabilistic, counteracting the advantages
that the entangled photon states might have. Here, we use a semiconductor
quantum dot to generate entangled multi-photon states in a deterministic
manner, using periodic timed excitation of a confined spin. This way
we entangle photons one-by-one at a rate which exceeds 300 MHz. We
use the resulting multi-photon state to demonstrate super-resolved
optical phase measurement. Our results open up a scalable way for
realizing genuine quantum enhanced super-sensitive measurements in
the near future.
\end{abstract}
\maketitle
\global\long\def\ket#1{\mathopen{|}#1\mathclose{\rangle}}%
\global\long\def\bra#1{\mathopen{\langle}#1\mathclose{|}}%
\global\long\def\im{\operatorname{Im}}%
\global\long\def\Tr{\operatorname{Tr}}%
\global\long\def\sgn{\operatorname{sign}}%

\section{Introduction}

When a light beam passes through a thin layer of transparent material,
it gains a phase shift relative to the same beam in vacuum. The shift
depends, in general, on the thickness of the layer, its refractive
index and its birefringence. Measuring the optical phase has, therefore,
numerous applications in science and technology, including microscopy,
lithography and displacement measurements, to name a few.

The precision in which such measurements can be performed is typically
limited to the shot-noise-limit (SNL) of $\text{\ensuremath{\Delta}}\theta_{\text{clas}}=1/\sqrt{N}$,
where $N$ is the total number of the detected beam photons. A possible
way to overcome this limit is to use entangled multi-photon states,
which can conceptually push the measurement precision towards the
Heisenberg limit of $1/N$ \citep{Lee2002,Giovannetti2011}.

A well known example is the N00N state \citep{Boto2000,Mitchell2004}.
Such a state of $N_{\text{ent}}$ photons can be expressed as $\left(\ket{N_{\text{ent}},0}+\ket{0,N_{\text{ent}}}\right)/\sqrt{2}$,
representing a superposition of all $N_{\text{ent}}$ photons in one
mode or all in another mode, with a well defined quantum mechanical
phase between the two. If one mode experiences a phase shift of $\theta$
relative to the other by passing a medium, the entangled multi-photon
state transfers to $\left(\ket{N_{\text{ent}},0}+e^{iN_{\text{ent}}\theta}\ket{0,N_{\text{ent}}}\right)/\sqrt{2}$.
The gained phase of $N_{\text{ent}}\theta$ can be accurately measured
using interferometry, for example, yielding a measure of $\theta$
with an error of $\Delta\theta_{\text{ent}}=1/N_{\text{ent}}$. Since
only a single mode emerging from a single source experiences the phase
shift, the N00N states also provide high spatial resolution when measuring
local phase shifts \citep{Dowling2008}. Unfortunately, generating
a N00N state is a very demanding and resource intensive task, thus
only N00N states with $N_{\text{ent}}=5$ photons have been reported
so far \citep{Afek879}.

The Greenberger-Horne-Zeilinger (GHZ) \citep{Walther2004} state is
yet another multi-photon entangled state that can be used for super-sensitive
phase measurements. It is expressed as $\bigl(\ket 0^{\otimes N_{\text{ent}}}+e^{i\alpha}\ket 1^{\otimes N_{\text{ent}}}\bigr)/\sqrt{2}$,
describing a superposition of $N_{\text{ent}}$ photons, all in state
$\ket 0$ or all in state $\ket 1$, with a well defined relative
phase $\alpha$ between the two cases. Similarly to the N00N case,
if one of the states experiences a phase shift of $\theta$ relative
to the other, $\theta$ can, in principle, be measured in the Heisenberg
accuracy limit.

In this work, we produce such a GHZ state where $\ket 0$ and $\ket 1$
are implemented in two orthogonal polarizations of the photons. GHZ
states have already been produced with up to 12 \citep{Zhong.12-photon.2018}
and 18 photonic qubits \citep{Wang2018} using spontaneous parametric
down converted (SPDC) light sources \citep{Pan2012}. Nevertheless,
these sources are probabilistic, and require inefficient post-selection
in order to create the GHZ states. In addition, the spatial resolution
that such GHZ states can provide, is relatively limited. This is because
the generated GHZ states occupy multiple spatial modes. These and
other requirements challenge the use of SPDC sources as suitable and
scalable sources for super-sensitive phase measurement applications.

Single photon sources with spontaneously generated entanglement, were
also considered recently for achieving super-sensitivity \citep{Motes.2015}.
Single photon sources based on semiconductor quantum dots (QDs) are
particularly bright and capable of deterministic production of single
\citep{E.Dekel.2000,Somaschi2016,Ding2016} and entangled \citep{Akopian2006,Young_2006,Muller2014,Roni.Winik.2017,Liu2019,Wang2019}
photons. Attempts to demonstrate phase super-sensitive measurements
were recently reported using entangled two-photon ($N_{\text{ent}}=2$)
states from a single QD \citep{Bennette2016,Muller2017}. Unfortunately,
these methods are intrinsically limited to low numbers of entangled
photons \citep{Olson2017,Zuen2017}.

Here, we demonstrate for the first time a new approach for achieving
super-sensitive optical phase measurement. This approach utilizes
semiconductor QDs to deterministically generate multi-photon, polarization-entangled,
GHZ states. We do it by periodic pulsed excitation of the QD, entangling
photons at a rate of 330 MHz. The number of photons that can be entangled
($N_{\text{ent}}$), this way is in principle unlimited. In addition,
the produced GHZ states occupy one spatial mode, providing the highest
spatial resolution possible. These advantages pave the way for building
a scalable method for performing super-sensitive measurements. We
describe below the experiment that demonstrates the concept and discuss
the conditions for achieving genuine super-sensitivity.

\section{Theoretical Background}

\subsection{Optical phase measurement with classical light}

\begin{figure}
\begin{centering}
\includegraphics[width=1\columnwidth]{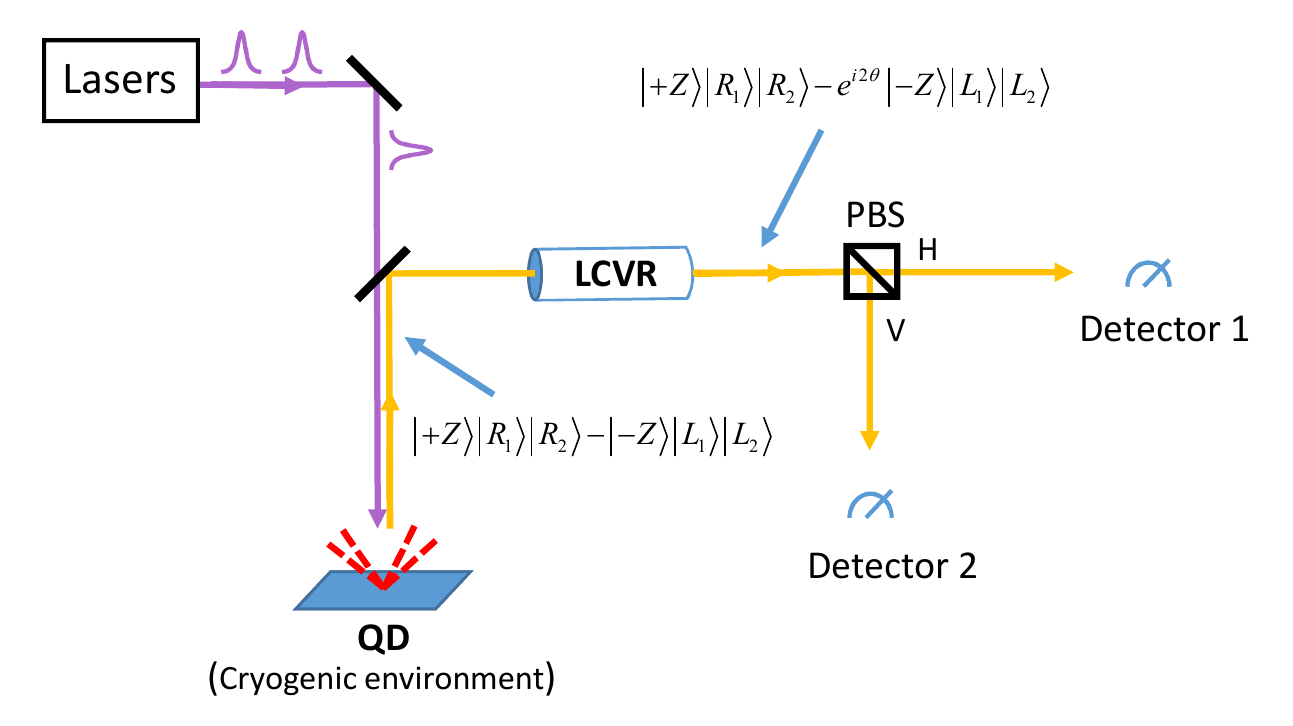}
\par\end{centering}
\caption{\label{fig:Experimental setup} Schematic and simplified description
of the experimental system. A sequence of pulses is applied to the
quantum dot (QD) at 76 MHz. The pulses deterministically generate
a string of photons, which are polarization entangled with the spin
of the dark exciton (DE) in the QD. The emitted photons pass through
a liquid crystal variable retarder (LCVR), which adds adjustable relative
phase difference between the two components of the light circular
polarizations R and L. The polarization of the photons is then projected
on two rectilinear polarizations H and V using a polarizing beam splitter
(PBS). Correlation events in which two or three clicks occur during
the same period are recorded by the time-tagging electronics.}
\end{figure}

Consider the experimental setup described in Fig.~\ref{fig:Experimental setup}.
We set a liquid crystal variable retarder (LCVR) to add a relative
phase of $\theta$ between left- and right-circularly polarized light
transmitted through the LCVRs. For example, rectilinear horizontally
polarized light $\ket H=\left(\ket R+\ket L\right)/\sqrt{2}$ accumulates
a phase of $\theta$ upon transmission through the LCVR to become
$\left(\ket R+e^{i\theta}\ket L\right)/\sqrt{2}$. A way to measure
the accumulated phase $\theta$ is to project the light on a polarizing
beam splitter (PBS). One measures then the degree of rectilinear polarization
at the output, which is given by $D_{\text{RP}}(\theta)=\frac{I_{H}(\theta)-I_{V}(\theta)}{I_{H}(\theta)+I_{V}(\theta)}$,
where $I_{H}(\theta)$ ($I_{V}(\theta)$) is the intensity of the
light transmitted (reflected) by the PBS. It is straightforward to
show that
\begin{equation}
\begin{aligned}I_{H(V)} & (\theta)\varpropto\frac{1}{2}[1\pm D_{\text{RP}}(\theta))]\\
D_{\text{RP}}(\theta) & =D_{\text{RP}}^{\text{S}}\cos(\theta)
\end{aligned}
\label{eq: Nq-GHZ-2}
\end{equation}
where $D_{\text{RP}}^{\text{S}}$ is the degree of rectilinear polarization
of the light source before the LCVR (ideally $D_{\text{RP}}^{\text{S}}=1$).

The best uncertainty in determining $\theta$ , $\Delta\theta$ is
therefore given by:
\begin{equation}
\begin{aligned}\Delta\theta & =\frac{\Delta D_{\text{RP}}}{\partial D_{\text{RP}}(\theta)/\partial\theta}\\
 & =\frac{\Delta D_{\text{RP}}}{D_{\text{RP}}^{S}|\partial\cos(\theta)/\partial\theta|_{\cos(\theta)=0}}=\frac{\Delta D_{\text{RP}}}{D_{\text{RP}}^{\text{S}}}
\end{aligned}
\label{eq: Nq-GHZ-1}
\end{equation}
where one chooses the angle $\theta$ such that $D_{\text{RP}}(\theta)$
almost vanishes, and its slope maximizes. Here $\Delta D_{\text{RP}}$
is the experimental uncertainty in measuring the degree of rectilinear
polarization after the LCVR, for $\theta$ close to such a point ($\theta\simeq\pi/2)$.
For classical light this uncertainty is given precisely by $1/\sqrt{N}$,
where $N$ is the total number of photons used for measuring $D_{\text{RP}}$.
It follows that $\Delta\theta_{\text{clas}}=\frac{1}{D_{\text{RP}}^{\text{S}}\sqrt{N}}$.

\subsection{Optical phase measurement with entangled light}

For non-classical light composed of $N/N_{\text{ent}}$ bunches of
$N_{\text{ent}}$ entangled photons in each bunch, forming a GHZ state,
$\bigl(\ket R^{\otimes N_{\text{ent}}}+\ket L^{\otimes N_{\text{ent}}}\bigr)/\sqrt{2}$,
the considerations are slightly different. This time, transmission
through the LCVR results in accumulated phase of $N_{\text{ent}}\theta$
between the left and right polarization components $\bigl(\ket R^{\otimes N_{\text{ent}}}+e^{iN_{\text{ent}}\theta}\ket L^{\otimes N_{\text{ent}}}\bigr)/\sqrt{2}$.
Measuring the degree of rectilinear polarization in this case allows
the determination of $\theta$ with higher accuracy. To see this,
one obtains, as before, (see Eq.~(\ref{eq: Nq-GHZ-1}))

\begin{equation}
\begin{aligned}I_{H(V)}^{N_{\text{ent}}}(\theta) & \varpropto\frac{1}{2}[1\pm D_{\text{RP}}^{N_{\text{ent}}}(\theta)]\\
D_{\text{RP}}^{N_{\text{ent}}}(\theta) & =D_{\text{RP}}^{\text{S},N_{\text{ent}}}\text{cos}(N_{\text{ent}}\theta)
\end{aligned}
\label{eq: Nq-GHZ-1-2}
\end{equation}
where $D_{\text{RP}}^{\text{S},N_{\text{ent}}}$ is the degree of
rectilinear polarization of the entangled light source. Substituting
this in the expression for $\Delta\theta$, recalling that in this
case the uncertainty in the measured polarization degree is given
by the number of bunches: $\Delta D_{\text{RP}}^{N_{\text{ent}}}=(N/N_{\text{ent}}$$)^{-\frac{1}{2}}$
yields

\begin{equation}
\begin{aligned}\Delta\theta_{N_{\text{ent}}} & =\frac{\Delta D_{\text{RP}}^{N_{\text{ent}}}}{D_{\text{RP}}^{\text{S},N_{\text{ent}}}|\partial\text{cos}(N_{\text{ent}}\theta)/\partial\theta|_{\text{cos}(N_{\text{ent}}\theta)=0}}\\
 & =\frac{1}{D_{\text{RP}}^{\text{S},N_{\text{ent}}}\sqrt{N}\sqrt{N_{\text{ent}}}}\,=\frac{\Delta\theta_{\text{clas}}}{\sqrt{N_{\text{ent}}}}
\end{aligned}
\label{eq: Nq-GHZ-1-1}
\end{equation}
which means that if the initial degree of rectilinear polarization
$D_{\text{RP}}^{\text{S},N_{\text{ent}}}$ of the entangled light
is the same as that of the classical light ($D_{\text{RP}}^{\text{S},N_{\text{ent}}}=D_{\text{RP}}^{\text{S}}=D_{\text{RP}}^{\text{S},1})$
and if all $N$ photons are detected, the sensitivity of the optical
phase measurement with entangled light is $\sqrt{N_{\text{ent}}}$
times better than that of the classical light.

Eq.~(\ref{eq: Nq-GHZ-1-1}) holds for the ideal case in which each
bunch of $N_{\text{ent}}$ photons is maximally entangled and the
efficiency of the photon detection, $\eta,$ is 1. In reality, however,
the situation is different \citep{Resch2007,Nagata2007}. The system
detection efficiency is limited, and therefore for a finite $\eta$
, the efficiency of detecting $N_{\text{ent}}$-photon events is given
by $\eta^{N_{\text{ent}}}$. This means that in order to reach genuine
super-sensitivity even with entangled light of only $N_{\text{ent}}=2$,
$\eta$ should exceed 0.71. For super-sensitivity which is order of
magnitude better than the classical limit, $N_{\text{ent}}$ should
be more than 100, and $\eta$ should be better than 98\% .

Another obstacle in reaching genuine super-sensitivity is the deviation
of the multi-photon entangled state from a pure state. Typically,
due to various decoherence processes in the state generation, adding
photons to the multi-photon state results in greater coherence loss.
This loss can often be described by a characteristic exponential decay
in the degree of rectilinear polarization $D_{\text{RP}}^{\text{S},N_{\text{ent}}}$
of the entangled light source, as the number of entangled photons
$N_{\text{ent}}$ increases:
\begin{align}
D_{\text{RP}}^{\text{S},N_{\text{ent}}} & =D_{\text{RP}}^{\text{S},1}\,e^{-\left(N_{\text{ent}}-1\right)/N_{\text{D}}}\label{eq: Nq-GHZ-1-1-1}
\end{align}
Here, $D_{\text{RP}}^{\text{S},1}$ is the degree of rectilinear polarization
of the classical light beam composed of single non-entangled photons
and $N_{\text{D}}$ is a characteristic polarization decay length
of the entangled photon string.

With this dependence, although the increase in the string length improves
the sensitivity as $\sqrt{N_{\text{ent}}}$, at the same time the
exponential decay of the $D_{\text{RP}}$ reduces it. For the ideal
case in which $D_{\text{RP}}^{\text{S},1}=1$ and $\eta=1$ it is
straightforward to show that for a given $N_{\text{D}}$ maximum sensitivity
is obtained when $N_{\text{ent}}=N_{\text{D}}/2$. With this at hand,
it follows that super-sensitivity which is about 10\% better than
the SNL can be achieved with $N_{\text{ent}}=2$ entangled photons
from an entangled light source with $N_{\text{D}}\geq4$. In order
to get super-sensitivity which is an order of magnitude better than
the SNL, bunches longer than $N_{\text{ent}}=270$ entangled photons
are required from a light source with $N_{\text{D}}\geq540$. The
limit in which $N_{\text{D}}\rightarrow\infty$; $D_{\text{RP}}^{\text{S},1}=1;$
and $\eta=1$ is called the Heisenberg limit.

\section{Experiment}

\subsection{The dark exciton as a photon entangler}

\begin{figure}
\begin{centering}
\includegraphics[width=1\columnwidth]{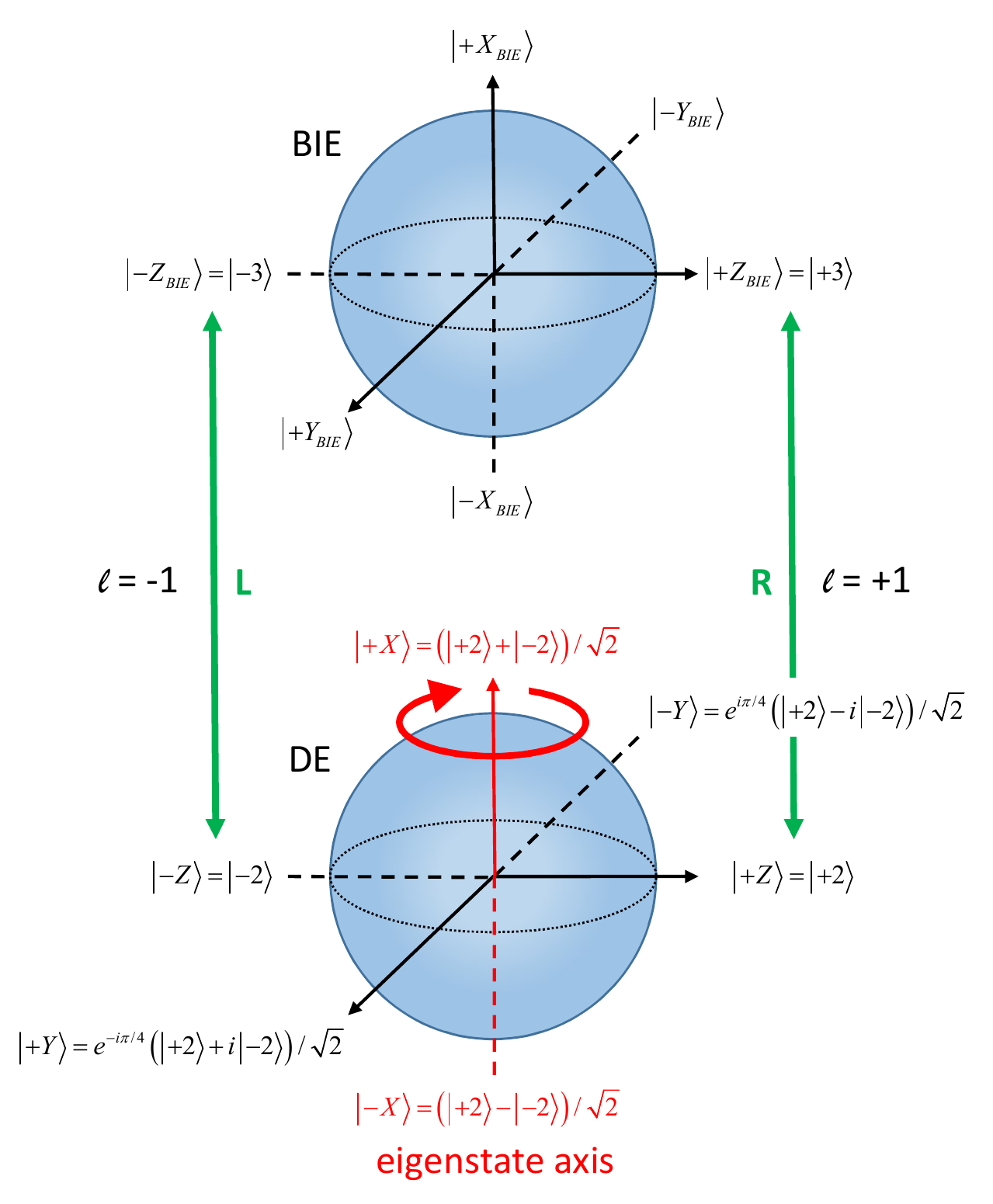}
\par\end{centering}
\caption{\label{fig:Bloch}The Bloch spheres describing the dark exciton (DE)
spin two level system and the biexciton (BIE). The degeneracy of the
qubits two eigenstates $\protect\ket{\pm X}$ is lifted by the exchange
interaction leading to precession around the $\hat{x}$-axis of any
coherent superposition of the qubit's eigenstates. The circularly
polarized optical transitions which connects between the two qubits
are marked by the green vertical arrows.}
\end{figure}

We use a QD to implement a scheme for deterministic generation of
a string of entangled photons \citep{cluster}. A QD-confined dark
exciton (DE), forms a physical two-level system, effectively acting
as a matter spin qubit (Fig.~\ref{fig:Bloch}) \citep{dark}.

Its two total spin ($2$) projections on the QD symmetry axis $\hat{z}$
form a basis, $\ket{\pm Z}=\ket{\pm2}$, for the DE qubit space. The
DE energy eigenstates are $\ket{\pm X}=\left(\ket{+Z}\pm\ket{-Z}\right)/\sqrt{2}$,
with an energy splitting $\Delta$$\varepsilon_{2}=1.5\,\text{\ensuremath{\mu}eV}$.
In the Bloch sphere representation, this splitting corresponds to
a coherent state-precession around the $\hat{x}$ axis, with a period
of $T_{\text{DE}}=h/\Delta\varepsilon_{2}\backsimeq3$ nsec \citep{dark}.
In addition to the DE, we use two states of a biexciton (BIE)—a bound
state of two excitons—whose total spin projections on the spatial
$\hat{z}$ axis are either $+3$ or $-3.$ The BIE eigenstates $\ket{\pm X_{\text{BIE}}}=\left(\ket{+Z_{\text{BIE}}}\pm\ket{-Z_{\text{BIE}}}\right)/\sqrt{2}$
are also non-degenerate, having precession period of $T_{\text{BIE}}=h/\Delta\varepsilon_{3}\backsimeq5$
nsec \citep{Cogan.Depolarization}. We denote these states by $\ket{\pm3}$.
The experimental protocol relies on the optical transition rules $\ket{+2}\longleftrightarrow\ket{+3}$
and $\ket{-2}\longleftrightarrow\ket{-3}$ through right hand $\ket R$
and left hand $\ket L$ circularly polarized photons, respectively
(see Fig.~\ref{fig:Bloch}).

\begin{figure}
\begin{centering}
\includegraphics[width=1\columnwidth]{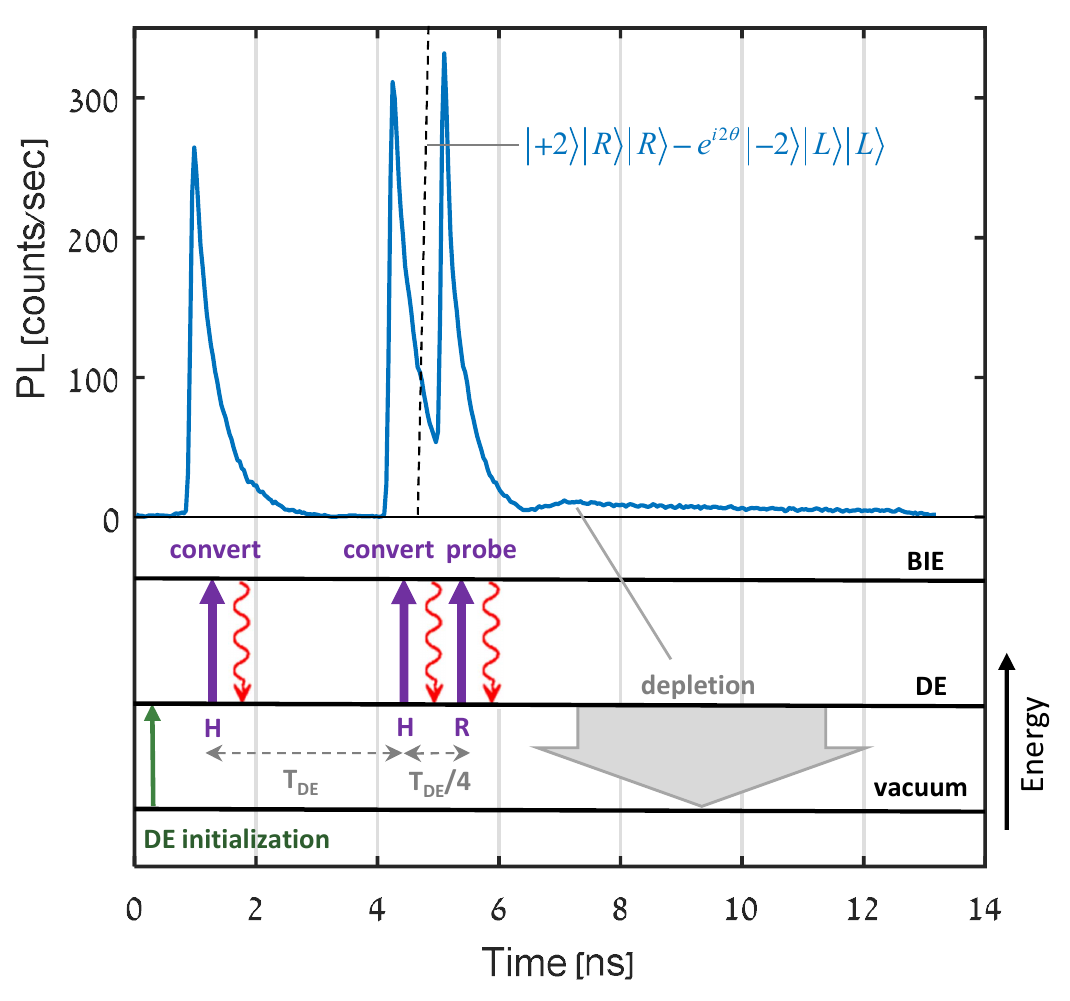}
\par\end{centering}
\caption{\label{fig:PL+scheme}Time-resolved photoluminscence (PL) signal (upper
panel) and the 76 MHz pulse sequence for generating the entangled
GHZ state (lower panel). Within 13 nsec the DE is initialized (green
upward arrow), converted $N_{\text{ent}}=2$ times to the BIE level
using rectilinearly polarized $\pi$ -pulses (purple arrows), separated
apart by the DE precession time. As a result, $N_{\text{ent}}$ photons
are emitted (curly downward red arrows), pass the LCVR and projected
on the rectilinear basis by the PBS. The last pulse is circularly
polarized, and timed quarter of a precession after the last pulse.
The detection of the resulted emitted photon projects the DE spin
on the $\protect\ket Y$ basis. The emitted photons are detected by
single photon detectors, as shown by the time resolved trace in the
upper panel. Detection of three-photon events during one period forms
a projection of the multiqubit $\protect\ket{\psi\left(\theta\right)_{\text{DE-1ph-2ph}}}$
GHZ state.}
\end{figure}

The pulse sequence for generating the $\ket{\text{GHZ}}$ state is
schematically described in the lower panel of Fig.~\ref{fig:PL+scheme}.
 It is executed at a rate of 76 MHz, corresponding to a time window
of \textasciitilde 13 nsec. Within each time window, a $\ket{\text{GHZ}}$
state is generated and used for the optical phase measurement.

First, we deterministically initialize the DE in its spin eigenstate
$\ket{\psi_{\text{DE}}^{\text{init}}}=\ket{-X}=\left(\ket{+2}-\ket{-2}\right)/\sqrt{2}$
using a short $\pi$-area picosecond pulse \citep{dark}. After the
initialization, we repeatedly apply a cycle containing three elements:
(i) a converting laser $\pi$-pulse, resonantly tuned to the DE-BIE
optical transition; (ii) subsequent radiative recombination of the
BIE, resulting in an emission of a photon entangled with the spin
of the DE which remains confined in the QD; and (iii) timed free precession
of the DE spin for one full period. In the first step of the cycle
(i), the pulse is horizontally polarized $\ket H=\left(\ket R+\ket L\right)/\sqrt{2}$
– an equal superposition of right- and left-hand circular polarizations.
It converts the DE state into the BIE state: $\ket{\psi_{\text{BIE}}}=\left(\ket{+3}-\ket{-3}\right)/\sqrt{2}$,
keeping the same relative phase between the $\pm$ spin state components.
In step (ii), radiative recombination of this BIE results in an entangled
state of the DE spin and the emitted photon polarization $\sqrt{2}\ket{\psi_{\text{DE-1ph}}}=\left(\ket Z\ket{R_{1}}-\ket{-Z}\ket{L_{1}}\right)$.
In step (iii), the DE completes full rotation around the Bloch $\hat{x}$
axis, thus returning to its original state after the BIE decays:
\begin{equation}
\xymatrix{ & \ket{+Z}\arc[dr]\\
\ket{-Y}\arc[ur] &  & \ket{+Y}\arc[dl]\\
 & \ket{-Z}\arc[ul]
}
\label{eq:precession}
\end{equation}
where we subdivided the evolution into increments of quarters of precession
period.

The sequence of steps (i)–(ii) forms one full cycle. Repeating the
cycle again results in a second photon, whose polarization state is
entangled with that of the first photon and the spin of the remaining
DE, yielding the tripartite GHZ state: 
\begin{equation}
\sqrt{2}\,\ket{\psi_{\text{DE-1ph-2ph}}}=\left(\ket Z\ket{R_{1}}\ket{R_{2}}-\ket{-Z}\ket{L_{1}}\ket{L_{2}}\right)\label{eq: 3q-GHZ}
\end{equation}
 In terms of the pulse sequence (Fig.~\ref{fig:PL+scheme}), this
state is generated with the emission of the second BIE photon following
the second converting pulse.

This cycle can be applied $N_{\text{ent}}$ times to generate an entangled
$N_{\text{ent}}+1$ GHZ state, containing $N_{\text{ent}}$ photons
and a DE. 
\begin{align}
\sqrt{2}\,\ket{\psi_{\text{DE-1ph-2ph-\ensuremath{\dots}-\ensuremath{N_{\text{ent}}}}}} & =\ket Z\ket{R_{1}}\ket{R_{2}}...\ket{R_{\text{\ensuremath{N_{\text{ent}}}}}}\nonumber \\
 & -\ket{-Z}\ket{L_{1}}\ket{L_{2}}...\ket{L_{N_{\text{ent}}}}\label{eq: Nq-GHZ}
\end{align}
When the emitted photons pass through the retarder of Fig.~\ref{fig:Experimental setup},
the state evolves to:
\begin{align}
\sqrt{2}\,\ket{\psi_{\text{DE-1ph-2ph-\ensuremath{\dots}-\ensuremath{N_{\text{ent}}}}}} & =\ket Z\ket{R_{1}}\ket{R_{2}}...\ket{R_{N_{\text{ent}}}}\nonumber \\
 & -e^{iN_{\text{ent}}\theta}\ket{-Z}\ket{L_{1}}\ket{L_{2}}...\ket{L_{N_{\text{ent}}}}\label{eq: Nq-GHZ-3}
\end{align}
For the conclusion of the experiment a last circularly polarized $\pi$-pulse,
quarter of a precession time after the previous pulse projects the
DE spin on the $\ket Y$ basis, when a photon is detected. The cycle
then ends in a \textasciitilde 7 nsec long optical pulse, which depletes
the QD and prepares it for the next cycle \citep{Emma.APL.2015}.

\begin{figure*}
\begin{centering}
\includegraphics[width=1\textwidth]{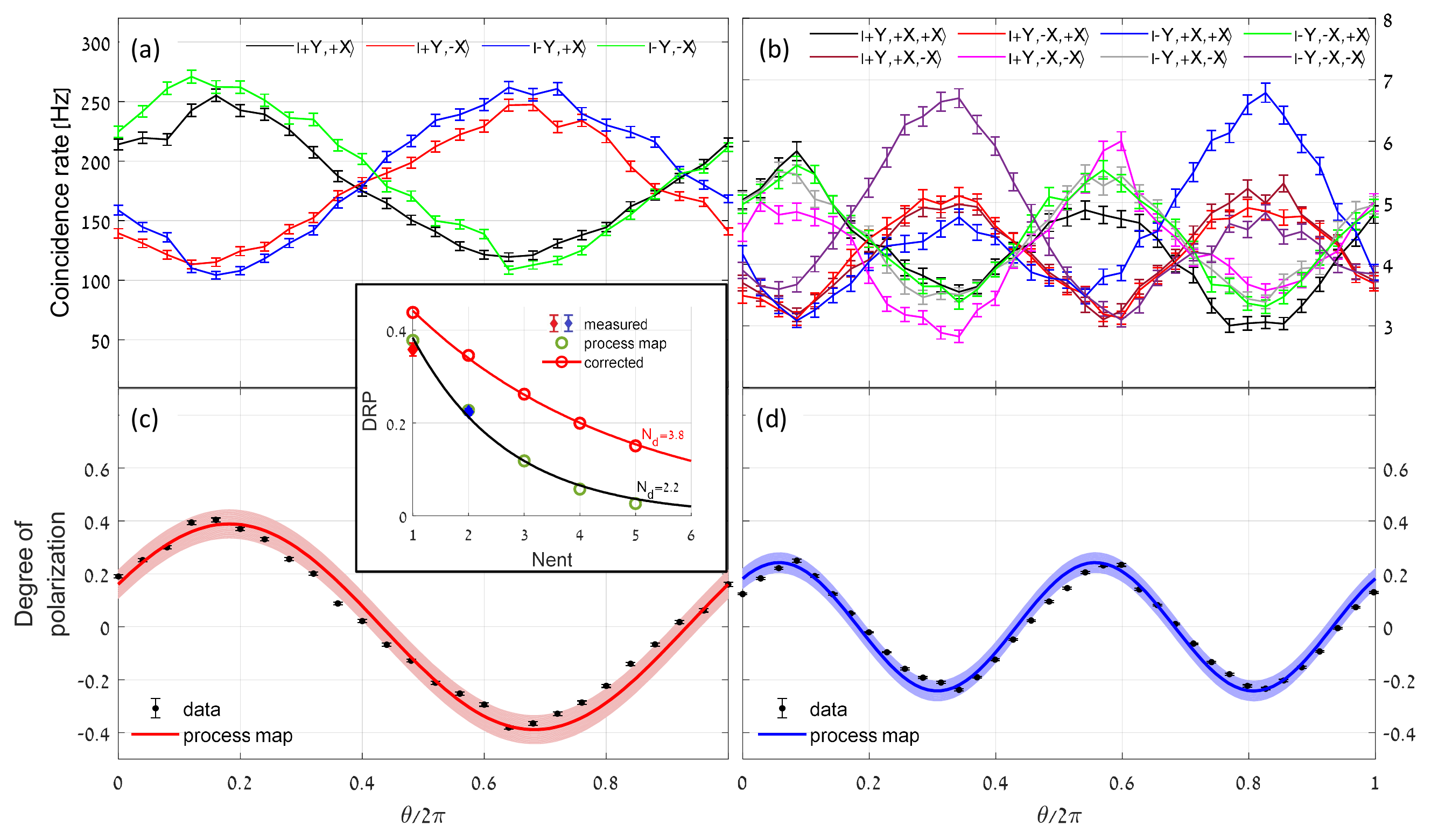}
\par\end{centering}
\caption[coincidence rates as a function of the LCVR phase $\theta$.]{\label{fig:summary_4}  Coincidence rates as a function of the LCVR
phase $\theta$. (a) Two-photon coincidence rate of all four $\protect\ket{\psi\left(\theta\right)_{\text{DE-1ph}}}$
state projections. Note that two have a positive $\cos\left(\theta\right)$
dependence and two negative $\cos\left(\theta\right)$ dependence
(see Eq.~(\ref{eq: Nq-GHZ-2})). (b) Three-photon coincidence rates
of all 8 $\protect\ket{\psi\left(\theta\right)_{\text{DE-1ph-2ph}}}$
state projections. Like in (a), four have positive $\cos\left(2\theta\right)$
dependence and four have negative $\cos\left(2\theta\right)$ dependence
(see Eq.~(\ref{eq: Nq-GHZ-1-2})). (c) Measured (symbols) and calculated
(solid red line) degree of rectilinear polarization ($D_{\text{RP}}^{1}(\theta))$.
(d) Measured (symbols) and calculated (solid blue line) $D_{\text{RP}}^{2}(\theta)$.
For the calculations of $D_{\text{RP}}^{1}(\theta)$ and $D_{\text{RP}}^{2}(\theta)$
we used the measured process map. The color matched shaded areas represent
the uncertainty in the calculations due to one standard deviation
uncertainty in the measured process map. The inset describes the measured
(symbols) and calculated from the process map (green circles) $D_{\text{RP}}^{\text{S},N_{\text{ent}}}$.
The black solid line describes characteristic exponential decay with
$N_{\text{D}}=2.2\pm0.2$, best fitted to measured and calculated
by the process map $D_{\text{RP}}^{\text{S},N_{\text{ent}}}$. The
red solid line describes best fitted exponential decay with $N_{\text{D}}=3.8\pm0.2$
to the calculated $D_{\text{RP}}^{\text{S},N_{\text{ent}}}$ corrected
for the depletion efficiency (red circles).}
\end{figure*}

\subsection{Calculating the multi-qubit quantum state using the repeated cycle's
process map}

The multi-qubit states that our method produces deviate from the pure
wavefunctions $\psi\left(\theta\right)_{\text{DE-1ph}}$ and $\psi\left(\theta\right)_{\text{DE-1ph-2ph-...Nent}}$,
described above. The proper way to describe our actual output state
is within the formalism of density matrices. We can calculate the
density matrix of the multi-qubit state that we produce by applying
repeatedly a linear transformation $\Phi$ to the initial state of
the DE. The transformation $\Phi$ is called a ``process map'' and
it describes the evolution of the system from $N_{\text{ent}}$ qubits
to $N_{\text{ent}}+1$ \citep{cluster}. For example, one can describe
the evolution of any initial DE state (spanned by a $2\times2$ density
matrix) when it is subjected to the excitation, photon emission and
full periodic precession of the DE, resulting in an entangled DE-photon
state (spanned by a $4\times4$ density matrix), by:
\begin{equation}
\rho_{\alpha\beta}^{\left(\text{DE+1ph}\right)}=\sum_{\mu}\Phi_{\alpha\beta}^{\mu}\rho_{\mu}^{\left(\text{DE}\right)}\label{eq: process}
\end{equation}
. Here, the density matrix elements are given in the Pauli basis,
such that $\hat{\rho}^{\left(\text{DE}\right)}=\sum_{\mu}\rho_{\mu}^{\left(\text{DE}\right)}\hat{\sigma}_{\mu}$,
$\hat{\rho}^{\left(\text{DE+1ph}\right)}=\sum_{\alpha\beta}\rho_{\alpha\beta}^{\left(\text{DE+1ph}\right)}\hat{\sigma}_{\alpha}\otimes\hat{\sigma}_{\beta}$,
where $\mu,\alpha,\beta\in0,1,2,3$. The process map has, therefore,
64 real parameters. To measure the process map, we first perform full
tomography of the initialized DE in six different initialization states
$\ket{\pm X},\ket{\pm Y},\ket{\pm Z}$ \citep{StateTomography.2020}.
Next, we apply to these states one cycle of our protocol and perform
full two-qubit tomography on the resulting entangled DE-photon states.
Finally, by solving a set of linear equations, the process map $\Phi$
is fully obtained \citep{cluster}. The fidelity of our measured map
to the ideal one, which describes an ideal two-qubit gate and no decoherence
at all, is 0.82.

Having the process map at hand, we apply it $N_{\text{ent}}$ times
to the measured initialization of the DE state, simulating the resulting
$\left(N_{\text{ent}}+1\right)$ GHZ state. Then, we add a phase of
$\cos(N_{\text{ent}}\theta)$ to the $\ket L\bra L$ component of
the density matrix relative to the $\ket R\bra R$ one, imitating
the action of the LCVRs on the transmitted photons. Finally we project
the simulated $N_{\text{ent}}+1$ qubits density matrix on the orthogonal
basis elements (photons on $\ket{\pm X}$ and spin on $\ket{\pm Y})$,
as done in the experiment.

\subsection{Experimental system}

A simplified version of the experimental system appears in Fig.~\ref{fig:Experimental setup}.
A sequence of laser pulses is launched on the QD, resulting in emission
of a string of single photons separated from each other by \textasciitilde 3
nsec. The photons are polarization entangled as explained above. By
passing through the LCVR an optical phase of $\theta$ is added to
$\ket L$ polarized photons relative to the $\ket R$ polarized ones.
Here we used the setup to produce $N_{\text{ent}}=1$ and $N_{\text{ent}}=2$
entangled spin-photon and entangled spin-photon-photon ($\ket{\text{GHZ}}$)
states, respectively. The photons are then projected using a standard
polarizing beam splitter (PBS) and detected using superconducting
single-photon detectors. In principle, one pair of detectors is enough
to perform the demonstration, provided that their recovery time is
shorter than the temporal separation of two sequential photons (3
nsec). In practice, since the recovery time of our detectors is longer
than that (\textasciitilde 20 nsec), we used two more detectors,
allowing us to measure up to four-photon correlations. We used a $\text{HydraHarp}^{\text{TM}}$
time-tagging device to record two- and three-photon coincidence events
for projection-measurements of the $N_{\text{ent}}=1$ and $N_{\text{ent}}=2$
cases, respectively. We recorded the coincidence rates while scanning
$\theta$ between 0 and $2\pi$. A coincidence event is registered
whenever two (or three) photons are detected within the same repetition
cycle of 13 nsec. The overall collection efficiency of our system,
is estimated as 1\%, thereby resulting in three-photon coincidence
rate of \textasciitilde 150 Hz.

\begin{figure}
\begin{centering}
\includegraphics[width=1\columnwidth]{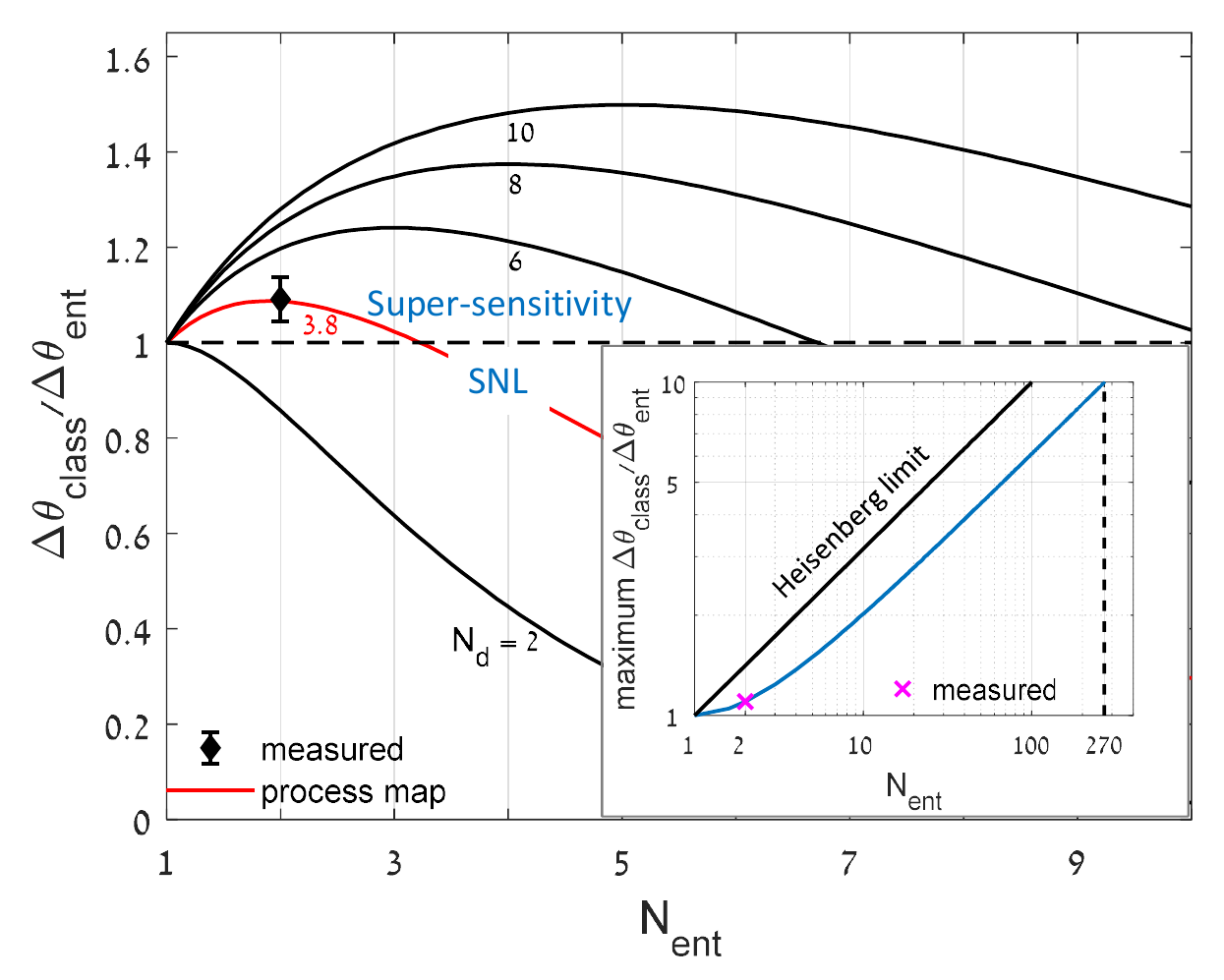}
\par\end{centering}
\caption{\label{fig:Visibility_decay} The enhancement in the optical phase
resolution relative to the SNL, as a function of the number of entangled
photons bunch length $N_{\text{ent}}$ for various characteristic
decay length $N_{\text{D}}$. The red curve represents the performance
of our device as deduced from its measured process map. The black
diamonds indicate the measured points, normalized for ideal initialization
of the DE. With better QD depletion, genuine super-sensitivity of
a few percents could be achieved. The inset shows maximum sensitivity
enhancement as a function of $N_{\text{ent}}=N_{\text{D}}$/2. An
order of magnitude better resolution requires bunches of $N_{\text{ent}}=270$
entangled photons from a source quality of $N_{\text{D}}=540$. The
solid diagonal line represents the case in which $N_{\text{D}}\rightarrow\infty,$
for which the Heisenberg limit is obtained.}
\end{figure}

\section{Results and discussion}

When a coincidence event is recorded, the data analysis proceeds as
follows: The detection of the last photon which results from the last
$\ket R$ ($\ket L$) circularly polarized excitation pulse, is used
to project the DE spin on the $\ket{+Y_{0}}$ ($\ket{-Y_{0}})$ base.
The preceding pulse(s), which result from $\ket H$ polarized excitation
pulse(s) are detected in either $\ket H$ or $\ket V$ polarization,
thereby projecting the detected photons on either the $\ket{+X}$
or $\ket{-X}$ basis states.

For the case of spin-photon entanglement $\ket{\psi\left(\theta\right)_{\text{DE-1ph}}}$
two-photon correlation measurements are used. In Fig.~\ref{fig:summary_4}a,
we present the measured coincidence rates as a function of $\theta$
for each one of the four possible projections. Two of them, $\ket{+Y_{0}}\ket{+X_{1}}$
and $\ket{-Y_{0}}\ket{-X_{1}}$, depend on $\theta$ through $A[1+D_{\text{RP}}^{\text{S},1}\cos(\theta)]$,
where $A$ is the average two-photon coincidence rate (see Eqs.~(\ref{eq: Nq-GHZ-2})
and (\ref{eq: Nq-GHZ-1})). We call this dependence a ``positive
$\cos(\theta)$'' dependence, referring to the plus sign coefficient
of $\cos(\theta)$. The other two projections, $\ket{+Y_{0}}\ket{-X_{1}}$
and $\ket{-Y_{0}}\ket{+X_{1}}$, have a ``negative $\cos(\theta)$''
dependence through $A[1-D_{\text{RP}}^{\text{S},1}\cos(\theta)]$.
Similarly, for the case of spin-photon-photon entanglement three-photon
correlation measurements are used. We then project the three-qubit
$\ket{\text{GHZ}}$ state, $\ket{\psi\left(\theta\right)_{\text{DE-1ph-2ph}}}$,
on 8 different possible polarization basis elements. Four of these
projections, namely: 
\begin{align*}
 & \ket{+Y_{0}}\ket{+X_{1}}\ket{+X_{2}},\,\ket{+Y_{0}}\ket{-X_{1}}\ket{-X_{2}},\\
 & \ket{-Y_{0}}\ket{-X_{1}}\ket{+X_{2}},\,\ket{-Y_{0}}\ket{+X_{1}}\ket{-X_{2}}
\end{align*}
 have positive $\cos(2\theta)$ dependence and four, obtained simply
by flipping all the signs in the expressions above, have negative
dependence (see Eq.~(\ref{eq: Nq-GHZ-1-2})). Fig.~\ref{fig:summary_4}b
presents the rate of three-photon coincidences as a function of $\theta$
for all these 8 projections.

The measured $D_{\text{RP}}^{N_{\text{ent}}}(\theta)$ for $N_{\text{ent}}=1$,
and 2 as deduced from Fig.~\ref{fig:summary_4}a and Fig.~\ref{fig:summary_4}b
are given by the data points in Fig.~\ref{fig:summary_4}c and in
Fig.~\ref{fig:summary_4}d respectively. As can be seen in these
figures the measured data points are indeed well described by functions
of the form $D_{\text{RP}}^{S,1}\cos\left(\theta\right)$ and $D_{\text{RP}}^{S,2}\cos\left(2\theta\right)$
respectively, as presented in the figures by the calculated solid
lines. 

The observed frequency doubling in the three-photon correlation events
as compared with the two-photon correlations, is termed ``super-resolution''.
It demonstrates the gain in the accuracy of the optical phase measurements,
resulting from the use of the $N_{\text{ent}}=2$ entangled photon
state.

We note here that in principle, the function $D_{\text{RP}}^{N_{\text{ent}}}(\theta)$
can be measured directly by our system, without measuring first coincidence
rates at various projections. This is because the sign dependence
of the measured DRP in a given bunch of $N_{\text{ent}}$ photons
can be deduced directly from the measurement results. In each individual
bunch measurement $i$, the degree of rectilinear polarization is
given by $D_{\text{RP}}^{i}=\frac{N_{H}^{i}-N_{V}^{i}}{N_{H}^{i}+N_{V}^{i}}$
where $N_{H}^{i}$ $(N_{V}^{i})$ is the number of $\ket H\,$($\ket V$)
polarized photons among the detected photons before the last detected
one in a given bunch. ($N_{H}^{i}+N_{V}^{i}=N_{\text{ent}}$ is equal
1(2) in Fig.~\ref{fig:summary_4}a (\ref{fig:summary_4}b)). The
sign of the $\cos(N_{\text{ent}}\theta)$ dependence is given by the
sign of the spin projection base $\ket{\pm Y_{0}^{i}}$ and the parity
of the number of photons detected in $\ket V$ polarization. The measured
degree of rectilinear polarization is therefore given by: 
\begin{equation}
D_{\text{RP}}=\sum_{i=1}^{N/N_{\text{ent}}}\sgn(\ket{\pm Y_{0}^{i}})(-1)^{N_{V}^{i}}D_{\text{RP}}^{i}\label{eq: DOP}
\end{equation}
 where $N$ is the total number of photons in the experiment.

In the inset to Fig.~\ref{fig:summary_4}c the measured (diamond-shape
marks), and calculated by the process map $D_{\text{RP}}^{\text{S},N_{\text{ent}}}$
(circle-shape marks) are presented. Using the measured and calculated
amplitudes in Eq.~(\ref{eq: Nq-GHZ-1-1-1}) one finds that the characteristic
decay of the $D_{\text{RP}}$ of our state is given by $N_{\text{D}}=2.2\pm0.2$.

We note that the $D_{\text{RP}}$ of the single photon beam that we
produced $D_{\text{RP}}^{\text{S},1}\backsimeq0.4$, is relatively
low. The reason for this is attributed to the limited efficiency (\textasciitilde 75\%)
by which we deplete the QD before the DE preparation \citep{Emma.APL.2015}.
This inefficient depletion, can be measured directly by the PL emission
intensity at the end of the depletion pulse (see Fig.~\ref{fig:PL+scheme}).
The limited depletion reduces both the fidelity of DE state preparation
and the fidelity of the DE spin projection, resulting in the measured
$D_{\text{RP}}^{\text{S},1}$ of 0.4 only. In fact, if one takes this
inefficiency into account corrects the initial state for it and applies
the process map on a fully depleted QD, the characteristic DRP decay
length becomes $N_{\text{D}}=3.8\pm0.2$, as can be seen in the inset
to Fig.~\ref{fig:summary_4}c. With this decay length genuine super-sensitivity
of a few percents can be achieved already with $N_{\text{ent}}=2$
entangled photons (and $\eta\geq0.71)$.

To see this we display in Fig.~\ref{fig:Visibility_decay} the calculated
enhancement of the optical phase resolution relative to the SNL, as
a function of $N_{\text{ent}}$. We display it for several sources
of varying quality, characterized by their DRP characteristic decay
lengths, $N_{\text{D}}$. As can be seen in Fig.~\ref{fig:Visibility_decay}
for a given source quality, an optimum is obtained if the number of
entangled photon used ($N_{\text{ent}}$) equals half of the characteristic
decay length. Using this condition we plot in the inset to Fig.~\ref{fig:Visibility_decay}
the enhancement in the optical phase measurement with respect to the
SNL as a function of the number of entangled photons in a bunch under
this condition ($N_{\text{ent}}=N_{D}/2$). The few percent super-resolution
that we achieved is represented by the data point in Fig.~\ref{fig:Visibility_decay}
and in its inset. The case in which all the photons in a given bunch
are maximally entangled ($N_{\text{D}}\rightarrow\infty$) is represented
in the inset by a dash line (Heisenberg limit).

In summary, we have demonstrated a novel way for achieving super-sensitivity
in optical phase measurement using deteministically prepared entangled
multi photon GHZ state. We outlined the required conditions for achieving
genuine super-sensitivity and showed that there are no conceptual
physical barriers which prevent achieving this long-desired technological
goal.

\begin{acknowledgments}
The support of the Israeli Science Foundation (ISF) and that of the
European Research Council (ERC) under the European Union’ s Horizon
2020 research and innovation program (Grant Agreement No. 695188)
are gratefully acknowledged.
\end{acknowledgments}

%


\begin{thebibliography}{31}%
\makeatletter
\providecommand \@ifxundefined [1]{%
 \@ifx{#1\undefined}
}%
\providecommand \@ifnum [1]{%
 \ifnum #1\expandafter \@firstoftwo
 \else \expandafter \@secondoftwo
 \fi
}%
\providecommand \@ifx [1]{%
 \ifx #1\expandafter \@firstoftwo
 \else \expandafter \@secondoftwo
 \fi
}%
\providecommand \natexlab [1]{#1}%
\providecommand \enquote  [1]{``#1''}%
\providecommand \bibnamefont  [1]{#1}%
\providecommand \bibfnamefont [1]{#1}%
\providecommand \citenamefont [1]{#1}%
\providecommand \href@noop [0]{\@secondoftwo}%
\providecommand \href [0]{\begingroup \@sanitize@url \@href}%
\providecommand \@href[1]{\@@startlink{#1}\@@href}%
\providecommand \@@href[1]{\endgroup#1\@@endlink}%
\providecommand \@sanitize@url [0]{\catcode `\\12\catcode `\$12\catcode
  `\&12\catcode `\#12\catcode `\^12\catcode `\_12\catcode `\%12\relax}%
\providecommand \@@startlink[1]{}%
\providecommand \@@endlink[0]{}%
\providecommand \url  [0]{\begingroup\@sanitize@url \@url }%
\providecommand \@url [1]{\endgroup\@href {#1}{\urlprefix }}%
\providecommand \urlprefix  [0]{URL }%
\providecommand \Eprint [0]{\href }%
\providecommand \doibase [0]{http://dx.doi.org/}%
\providecommand \selectlanguage [0]{\@gobble}%
\providecommand \bibinfo  [0]{\@secondoftwo}%
\providecommand \bibfield  [0]{\@secondoftwo}%
\providecommand \translation [1]{[#1]}%
\providecommand \BibitemOpen [0]{}%
\providecommand \bibitemStop [0]{}%
\providecommand \bibitemNoStop [0]{.\EOS\space}%
\providecommand \EOS [0]{\spacefactor3000\relax}%
\providecommand \BibitemShut  [1]{\csname bibitem#1\endcsname}%
\let\auto@bib@innerbib\@empty
\bibitem [{\citenamefont {Lee}, \citenamefont {Kok},\ and\ \citenamefont
  {Dowling}(2002)}]{Lee2002}%
  \BibitemOpen
  \bibfield  {author} {\bibinfo {author} {\bibfnamefont {H.}~\bibnamefont
  {Lee}}, \bibinfo {author} {\bibfnamefont {P.}~\bibnamefont {Kok}}, \ and\
  \bibinfo {author} {\bibfnamefont {J.}~\bibnamefont {Dowling}},\ }\href@noop
  {} {\bibfield  {journal} {\bibinfo  {journal} {Journal of Modern Optics}\
  }\textbf {\bibinfo {volume} {49}},\ \bibinfo {pages} {2325} (\bibinfo {year}
  {2002})}\BibitemShut {NoStop}%
\bibitem [{\citenamefont {Giovannetti}, \citenamefont {Lloyd},\ and\
  \citenamefont {Maccone}(2011)}]{Giovannetti2011}%
  \BibitemOpen
  \bibfield  {author} {\bibinfo {author} {\bibfnamefont {V.}~\bibnamefont
  {Giovannetti}}, \bibinfo {author} {\bibfnamefont {S.}~\bibnamefont {Lloyd}},
  \ and\ \bibinfo {author} {\bibfnamefont {L.}~\bibnamefont {Maccone}},\ }\href
  {https://doi.org/10.1038/nphoton.2011.35} {\bibfield  {journal} {\bibinfo
  {journal} {Nature Photonics}\ }\textbf {\bibinfo {volume} {5}},\ \bibinfo
  {pages} {222 EP } (\bibinfo {year} {2011})}\BibitemShut {NoStop}%
\bibitem [{\citenamefont {Boto}\ \emph {et~al.}(1999)\citenamefont {Boto},
  \citenamefont {Kok}, \citenamefont {Abrams}, \citenamefont {Braunstein},
  \citenamefont {Williams},\ and\ \citenamefont {Dowling}}]{Boto2000}%
  \BibitemOpen
  \bibfield  {author} {\bibinfo {author} {\bibfnamefont {A.}~\bibnamefont
  {Boto}}, \bibinfo {author} {\bibfnamefont {P.}~\bibnamefont {Kok}}, \bibinfo
  {author} {\bibfnamefont {D.}~\bibnamefont {Abrams}}, \bibinfo {author}
  {\bibfnamefont {S.}~\bibnamefont {Braunstein}}, \bibinfo {author}
  {\bibfnamefont {C.}~\bibnamefont {Williams}}, \ and\ \bibinfo {author}
  {\bibfnamefont {J.}~\bibnamefont {Dowling}},\ }\href
  {https://link.aps.org/doi/10.1103/PhysRevLett.85.2733} {\bibfield  {journal}
  {\bibinfo  {journal} {Physical Review Letters}\ }\textbf {\bibinfo {volume}
  {85}},\ \bibinfo {pages} {2733} (\bibinfo {year} {1999})}\BibitemShut
  {NoStop}%
\bibitem [{\citenamefont {Mitchell}, \citenamefont {Lundeen},\ and\
  \citenamefont {Steinberg}(2004)}]{Mitchell2004}%
  \BibitemOpen
  \bibfield  {author} {\bibinfo {author} {\bibfnamefont {M.~W.}\ \bibnamefont
  {Mitchell}}, \bibinfo {author} {\bibfnamefont {J.~S.}\ \bibnamefont
  {Lundeen}}, \ and\ \bibinfo {author} {\bibfnamefont {A.~M.}\ \bibnamefont
  {Steinberg}},\ }\href {\doibase 10.1038/nature02493} {\bibfield  {journal}
  {\bibinfo  {journal} {Nature}\ }\textbf {\bibinfo {volume} {429}},\ \bibinfo
  {pages} {161} (\bibinfo {year} {2004})}\BibitemShut {NoStop}%
\bibitem [{\citenamefont {Dowling}(2008)}]{Dowling2008}%
  \BibitemOpen
  \bibfield  {author} {\bibinfo {author} {\bibfnamefont {J.~P.}\ \bibnamefont
  {Dowling}},\ }\href {\doibase 10.1080/00107510802091298} {\bibfield
  {journal} {\bibinfo  {journal} {Contemporary Physics}\ }\textbf {\bibinfo
  {volume} {49}},\ \bibinfo {pages} {125} (\bibinfo {year} {2008})}\BibitemShut
  {NoStop}%
\bibitem [{\citenamefont {Afek}, \citenamefont {Ambar},\ and\ \citenamefont
  {Silberberg}(2010)}]{Afek879}%
  \BibitemOpen
  \bibfield  {author} {\bibinfo {author} {\bibfnamefont {I.}~\bibnamefont
  {Afek}}, \bibinfo {author} {\bibfnamefont {O.}~\bibnamefont {Ambar}}, \ and\
  \bibinfo {author} {\bibfnamefont {Y.}~\bibnamefont {Silberberg}},\ }\href
  {\doibase 10.1126/science.1188172} {\bibfield  {journal} {\bibinfo  {journal}
  {Science}\ }\textbf {\bibinfo {volume} {328}},\ \bibinfo {pages} {879}
  (\bibinfo {year} {2010})}\BibitemShut {NoStop}%
\bibitem [{\citenamefont {Walther}\ \emph {et~al.}(2004)\citenamefont
  {Walther}, \citenamefont {Pan}, \citenamefont {Aspelmeyer}, \citenamefont
  {Ursin}, \citenamefont {Gasparoni},\ and\ \citenamefont
  {Zeilinger}}]{Walther2004}%
  \BibitemOpen
  \bibfield  {author} {\bibinfo {author} {\bibfnamefont {P.}~\bibnamefont
  {Walther}}, \bibinfo {author} {\bibfnamefont {J.-W.}\ \bibnamefont {Pan}},
  \bibinfo {author} {\bibfnamefont {M.}~\bibnamefont {Aspelmeyer}}, \bibinfo
  {author} {\bibfnamefont {R.}~\bibnamefont {Ursin}}, \bibinfo {author}
  {\bibfnamefont {S.}~\bibnamefont {Gasparoni}}, \ and\ \bibinfo {author}
  {\bibfnamefont {A.}~\bibnamefont {Zeilinger}},\ }\href {\doibase
  10.1038/nature02552} {\bibfield  {journal} {\bibinfo  {journal} {Nature}\
  }\textbf {\bibinfo {volume} {429}},\ \bibinfo {pages} {158} (\bibinfo {year}
  {2004})}\BibitemShut {NoStop}%
\bibitem [{\citenamefont {Zhong}\ \emph {et~al.}(2018)\citenamefont {Zhong},
  \citenamefont {Li}, \citenamefont {Li}, \citenamefont {Peng}, \citenamefont
  {Su}, \citenamefont {Hu}, \citenamefont {He}, \citenamefont {Ding},
  \citenamefont {Zhang}, \citenamefont {Li}, \citenamefont {Zhang},
  \citenamefont {Wang}, \citenamefont {You}, \citenamefont {Wang},
  \citenamefont {Jiang}, \citenamefont {Li}, \citenamefont {Chen},
  \citenamefont {Liu}, \citenamefont {Lu},\ and\ \citenamefont
  {Pan}}]{Zhong.12-photon.2018}%
  \BibitemOpen
  \bibfield  {author} {\bibinfo {author} {\bibfnamefont {H.-S.}\ \bibnamefont
  {Zhong}}, \bibinfo {author} {\bibfnamefont {Y.}~\bibnamefont {Li}}, \bibinfo
  {author} {\bibfnamefont {W.}~\bibnamefont {Li}}, \bibinfo {author}
  {\bibfnamefont {L.-C.}\ \bibnamefont {Peng}}, \bibinfo {author}
  {\bibfnamefont {Z.-E.}\ \bibnamefont {Su}}, \bibinfo {author} {\bibfnamefont
  {Y.}~\bibnamefont {Hu}}, \bibinfo {author} {\bibfnamefont {Y.-M.}\
  \bibnamefont {He}}, \bibinfo {author} {\bibfnamefont {X.}~\bibnamefont
  {Ding}}, \bibinfo {author} {\bibfnamefont {W.}~\bibnamefont {Zhang}},
  \bibinfo {author} {\bibfnamefont {H.}~\bibnamefont {Li}}, \bibinfo {author}
  {\bibfnamefont {L.}~\bibnamefont {Zhang}}, \bibinfo {author} {\bibfnamefont
  {Z.}~\bibnamefont {Wang}}, \bibinfo {author} {\bibfnamefont {L.}~\bibnamefont
  {You}}, \bibinfo {author} {\bibfnamefont {X.-L.}\ \bibnamefont {Wang}},
  \bibinfo {author} {\bibfnamefont {X.}~\bibnamefont {Jiang}}, \bibinfo
  {author} {\bibfnamefont {L.}~\bibnamefont {Li}}, \bibinfo {author}
  {\bibfnamefont {Y.-A.}\ \bibnamefont {Chen}}, \bibinfo {author}
  {\bibfnamefont {N.-L.}\ \bibnamefont {Liu}}, \bibinfo {author} {\bibfnamefont
  {C.-Y.}\ \bibnamefont {Lu}}, \ and\ \bibinfo {author} {\bibfnamefont {J.-W.}\
  \bibnamefont {Pan}},\ }\href {\doibase 10.1103/PhysRevLett.121.250505}
  {\bibfield  {journal} {\bibinfo  {journal} {Phys. Rev. Lett.}\ }\textbf
  {\bibinfo {volume} {121}},\ \bibinfo {pages} {250505} (\bibinfo {year}
  {2018})}\BibitemShut {NoStop}%
\bibitem [{\citenamefont {Wang}\ \emph {et~al.}(2018)\citenamefont {Wang},
  \citenamefont {Luo}, \citenamefont {Huang}, \citenamefont {Chen},
  \citenamefont {Su}, \citenamefont {Liu}, \citenamefont {Chen}, \citenamefont
  {Li}, \citenamefont {Fang}, \citenamefont {Jiang}, \citenamefont {Zhang},
  \citenamefont {Li}, \citenamefont {Liu}, \citenamefont {Lu},\ and\
  \citenamefont {Pan}}]{Wang2018}%
  \BibitemOpen
  \bibfield  {author} {\bibinfo {author} {\bibfnamefont {X.-L.}\ \bibnamefont
  {Wang}}, \bibinfo {author} {\bibfnamefont {Y.-H.}\ \bibnamefont {Luo}},
  \bibinfo {author} {\bibfnamefont {H.-L.}\ \bibnamefont {Huang}}, \bibinfo
  {author} {\bibfnamefont {M.-C.}\ \bibnamefont {Chen}}, \bibinfo {author}
  {\bibfnamefont {Z.-E.}\ \bibnamefont {Su}}, \bibinfo {author} {\bibfnamefont
  {C.}~\bibnamefont {Liu}}, \bibinfo {author} {\bibfnamefont {C.}~\bibnamefont
  {Chen}}, \bibinfo {author} {\bibfnamefont {W.}~\bibnamefont {Li}}, \bibinfo
  {author} {\bibfnamefont {Y.-Q.}\ \bibnamefont {Fang}}, \bibinfo {author}
  {\bibfnamefont {X.}~\bibnamefont {Jiang}}, \bibinfo {author} {\bibfnamefont
  {J.}~\bibnamefont {Zhang}}, \bibinfo {author} {\bibfnamefont
  {L.}~\bibnamefont {Li}}, \bibinfo {author} {\bibfnamefont {N.-L.}\
  \bibnamefont {Liu}}, \bibinfo {author} {\bibfnamefont {C.-Y.}\ \bibnamefont
  {Lu}}, \ and\ \bibinfo {author} {\bibfnamefont {J.-W.}\ \bibnamefont {Pan}},\
  }\href {\doibase 10.1103/PhysRevLett.120.260502} {\bibfield  {journal}
  {\bibinfo  {journal} {Phys. Rev. Lett.}\ }\textbf {\bibinfo {volume} {120}},\
  \bibinfo {pages} {260502} (\bibinfo {year} {2018})}\BibitemShut {NoStop}%
\bibitem [{\citenamefont {Pan}\ \emph {et~al.}(2008)\citenamefont {Pan},
  \citenamefont {Chen}, \citenamefont {Lu}, \citenamefont {Weinfurter},
  \citenamefont {Zeilinger},\ and\ \citenamefont {Zukowski}}]{Pan2012}%
  \BibitemOpen
  \bibfield  {author} {\bibinfo {author} {\bibfnamefont {J.-W.}\ \bibnamefont
  {Pan}}, \bibinfo {author} {\bibfnamefont {T.}~\bibnamefont {Chen}}, \bibinfo
  {author} {\bibfnamefont {C.-Y.}\ \bibnamefont {Lu}}, \bibinfo {author}
  {\bibfnamefont {H.}~\bibnamefont {Weinfurter}}, \bibinfo {author}
  {\bibfnamefont {A.}~\bibnamefont {Zeilinger}}, \ and\ \bibinfo {author}
  {\bibfnamefont {M.}~\bibnamefont {Zukowski}},\ }\href
  {https://link.aps.org/doi/10.1103/RevModPhys.84.777} {\bibfield  {journal}
  {\bibinfo  {journal} {Reviews of Modern Physics}\ }\textbf {\bibinfo {volume}
  {84}},\ \bibinfo {pages} {777} (\bibinfo {year} {2008})}\BibitemShut
  {NoStop}%
\bibitem [{\citenamefont {Motes}\ \emph {et~al.}(2015)\citenamefont {Motes},
  \citenamefont {Olson}, \citenamefont {Rabeaux}, \citenamefont {Dowling},
  \citenamefont {Olson},\ and\ \citenamefont {Rohde}}]{Motes.2015}%
  \BibitemOpen
  \bibfield  {author} {\bibinfo {author} {\bibfnamefont {K.~R.}\ \bibnamefont
  {Motes}}, \bibinfo {author} {\bibfnamefont {J.~P.}\ \bibnamefont {Olson}},
  \bibinfo {author} {\bibfnamefont {E.~J.}\ \bibnamefont {Rabeaux}}, \bibinfo
  {author} {\bibfnamefont {J.~P.}\ \bibnamefont {Dowling}}, \bibinfo {author}
  {\bibfnamefont {S.~J.}\ \bibnamefont {Olson}}, \ and\ \bibinfo {author}
  {\bibfnamefont {P.~P.}\ \bibnamefont {Rohde}},\ }\href {\doibase
  10.1103/PhysRevLett.114.170802} {\bibfield  {journal} {\bibinfo  {journal}
  {Phys. Rev. Lett.}\ }\textbf {\bibinfo {volume} {114}},\ \bibinfo {pages}
  {170802} (\bibinfo {year} {2015})}\BibitemShut {NoStop}%
\bibitem [{\citenamefont {Dekel}\ \emph {et~al.}(2000)\citenamefont {Dekel},
  \citenamefont {Gershoni}, \citenamefont {Ehrenfreund}, \citenamefont
  {Garcia},\ and\ \citenamefont {Petroff}}]{E.Dekel.2000}%
  \BibitemOpen
  \bibfield  {author} {\bibinfo {author} {\bibfnamefont {E.}~\bibnamefont
  {Dekel}}, \bibinfo {author} {\bibfnamefont {D.}~\bibnamefont {Gershoni}},
  \bibinfo {author} {\bibfnamefont {E.}~\bibnamefont {Ehrenfreund}}, \bibinfo
  {author} {\bibfnamefont {J.~M.}\ \bibnamefont {Garcia}}, \ and\ \bibinfo
  {author} {\bibfnamefont {P.~M.}\ \bibnamefont {Petroff}},\ }\href {\doibase
  10.1103/PhysRevB.61.11009} {\bibfield  {journal} {\bibinfo  {journal} {Phys.
  Rev. B}\ }\textbf {\bibinfo {volume} {61}},\ \bibinfo {pages} {11009}
  (\bibinfo {year} {2000})}\BibitemShut {NoStop}%
\bibitem [{\citenamefont {Somaschi}\ \emph {et~al.}(2016)\citenamefont
  {Somaschi}, \citenamefont {Giesz}, \citenamefont {De~Santis}, \citenamefont
  {Loredo}, \citenamefont {Almeida}, \citenamefont {Hornecker}, \citenamefont
  {Portalupi}, \citenamefont {Grange}, \citenamefont {Ant{\'o}n}, \citenamefont
  {Demory}, \citenamefont {G{\'o}mez}, \citenamefont {Sagnes}, \citenamefont
  {Lanzillotti-Kimura}, \citenamefont {Lema{\'i}tre}, \citenamefont {Auffeves},
  \citenamefont {White}, \citenamefont {Lanco},\ and\ \citenamefont
  {Senellart}}]{Somaschi2016}%
  \BibitemOpen
  \bibfield  {author} {\bibinfo {author} {\bibfnamefont {N.}~\bibnamefont
  {Somaschi}}, \bibinfo {author} {\bibfnamefont {V.}~\bibnamefont {Giesz}},
  \bibinfo {author} {\bibfnamefont {L.}~\bibnamefont {De~Santis}}, \bibinfo
  {author} {\bibfnamefont {J.~C.}\ \bibnamefont {Loredo}}, \bibinfo {author}
  {\bibfnamefont {M.~P.}\ \bibnamefont {Almeida}}, \bibinfo {author}
  {\bibfnamefont {G.}~\bibnamefont {Hornecker}}, \bibinfo {author}
  {\bibfnamefont {S.~L.}\ \bibnamefont {Portalupi}}, \bibinfo {author}
  {\bibfnamefont {T.}~\bibnamefont {Grange}}, \bibinfo {author} {\bibfnamefont
  {C.}~\bibnamefont {Ant{\'o}n}}, \bibinfo {author} {\bibfnamefont
  {J.}~\bibnamefont {Demory}}, \bibinfo {author} {\bibfnamefont
  {C.}~\bibnamefont {G{\'o}mez}}, \bibinfo {author} {\bibfnamefont
  {I.}~\bibnamefont {Sagnes}}, \bibinfo {author} {\bibfnamefont {N.~D.}\
  \bibnamefont {Lanzillotti-Kimura}}, \bibinfo {author} {\bibfnamefont
  {A.}~\bibnamefont {Lema{\'i}tre}}, \bibinfo {author} {\bibfnamefont
  {A.}~\bibnamefont {Auffeves}}, \bibinfo {author} {\bibfnamefont {A.~G.}\
  \bibnamefont {White}}, \bibinfo {author} {\bibfnamefont {L.}~\bibnamefont
  {Lanco}}, \ and\ \bibinfo {author} {\bibfnamefont {P.}~\bibnamefont
  {Senellart}},\ }\href {https://doi.org/10.1038/nphoton.2016.23} {\bibfield
  {journal} {\bibinfo  {journal} {Nature Photonics}\ }\textbf {\bibinfo
  {volume} {10}},\ \bibinfo {pages} {340 EP } (\bibinfo {year}
  {2016})}\BibitemShut {NoStop}%
\bibitem [{\citenamefont {Ding}\ \emph {et~al.}(2016)\citenamefont {Ding},
  \citenamefont {He}, \citenamefont {Duan}, \citenamefont {Gregersen},
  \citenamefont {Chen}, \citenamefont {Unsleber}, \citenamefont {Maier},
  \citenamefont {Schneider}, \citenamefont {Kamp}, \citenamefont {H\"ofling},
  \citenamefont {Lu},\ and\ \citenamefont {Pan}}]{Ding2016}%
  \BibitemOpen
  \bibfield  {author} {\bibinfo {author} {\bibfnamefont {X.}~\bibnamefont
  {Ding}}, \bibinfo {author} {\bibfnamefont {Y.}~\bibnamefont {He}}, \bibinfo
  {author} {\bibfnamefont {Z.-C.}\ \bibnamefont {Duan}}, \bibinfo {author}
  {\bibfnamefont {N.}~\bibnamefont {Gregersen}}, \bibinfo {author}
  {\bibfnamefont {M.-C.}\ \bibnamefont {Chen}}, \bibinfo {author}
  {\bibfnamefont {S.}~\bibnamefont {Unsleber}}, \bibinfo {author}
  {\bibfnamefont {S.}~\bibnamefont {Maier}}, \bibinfo {author} {\bibfnamefont
  {C.}~\bibnamefont {Schneider}}, \bibinfo {author} {\bibfnamefont
  {M.}~\bibnamefont {Kamp}}, \bibinfo {author} {\bibfnamefont {S.}~\bibnamefont
  {H\"ofling}}, \bibinfo {author} {\bibfnamefont {C.-Y.}\ \bibnamefont {Lu}}, \
  and\ \bibinfo {author} {\bibfnamefont {J.-W.}\ \bibnamefont {Pan}},\ }\href
  {\doibase 10.1103/PhysRevLett.116.020401} {\bibfield  {journal} {\bibinfo
  {journal} {Phys. Rev. Lett.}\ }\textbf {\bibinfo {volume} {116}},\ \bibinfo
  {pages} {020401} (\bibinfo {year} {2016})}\BibitemShut {NoStop}%
\bibitem [{\citenamefont {Akopian}\ \emph {et~al.}(2006)\citenamefont
  {Akopian}, \citenamefont {Lindner}, \citenamefont {Poem}, \citenamefont
  {Berlatzky}, \citenamefont {Avron}, \citenamefont {Gershoni}, \citenamefont
  {Gerardot},\ and\ \citenamefont {Petroff}}]{Akopian2006}%
  \BibitemOpen
  \bibfield  {author} {\bibinfo {author} {\bibfnamefont {N.}~\bibnamefont
  {Akopian}}, \bibinfo {author} {\bibfnamefont {N.~H.}\ \bibnamefont
  {Lindner}}, \bibinfo {author} {\bibfnamefont {E.}~\bibnamefont {Poem}},
  \bibinfo {author} {\bibfnamefont {Y.}~\bibnamefont {Berlatzky}}, \bibinfo
  {author} {\bibfnamefont {J.}~\bibnamefont {Avron}}, \bibinfo {author}
  {\bibfnamefont {D.}~\bibnamefont {Gershoni}}, \bibinfo {author}
  {\bibfnamefont {B.~D.}\ \bibnamefont {Gerardot}}, \ and\ \bibinfo {author}
  {\bibfnamefont {P.~M.}\ \bibnamefont {Petroff}},\ }\href {\doibase
  10.1103/PhysRevLett.96.130501} {\bibfield  {journal} {\bibinfo  {journal}
  {Phys. Rev. Lett.}\ }\textbf {\bibinfo {volume} {96}},\ \bibinfo {pages}
  {130501} (\bibinfo {year} {2006})}\BibitemShut {NoStop}%
\bibitem [{\citenamefont {Young}\ \emph {et~al.}(2006)\citenamefont {Young},
  \citenamefont {Stevenson}, \citenamefont {Atkinson}, \citenamefont {Cooper},
  \citenamefont {Ritchie},\ and\ \citenamefont {Shields}}]{Young_2006}%
  \BibitemOpen
  \bibfield  {author} {\bibinfo {author} {\bibfnamefont {R.~J.}\ \bibnamefont
  {Young}}, \bibinfo {author} {\bibfnamefont {R.~M.}\ \bibnamefont
  {Stevenson}}, \bibinfo {author} {\bibfnamefont {P.}~\bibnamefont {Atkinson}},
  \bibinfo {author} {\bibfnamefont {K.}~\bibnamefont {Cooper}}, \bibinfo
  {author} {\bibfnamefont {D.~A.}\ \bibnamefont {Ritchie}}, \ and\ \bibinfo
  {author} {\bibfnamefont {A.~J.}\ \bibnamefont {Shields}},\ }\href {\doibase
  10.1088/1367-2630/8/2/029} {\bibfield  {journal} {\bibinfo  {journal} {New
  Journal of Physics}\ }\textbf {\bibinfo {volume} {8}},\ \bibinfo {pages} {29}
  (\bibinfo {year} {2006})}\BibitemShut {NoStop}%
\bibitem [{\citenamefont {M{\"u}ller}\ \emph {et~al.}(2014)\citenamefont
  {M{\"u}ller}, \citenamefont {Bounouar}, \citenamefont {J{\"o}ns},
  \citenamefont {Gl{\"a}ssl},\ and\ \citenamefont {Michler}}]{Muller2014}%
  \BibitemOpen
  \bibfield  {author} {\bibinfo {author} {\bibfnamefont {M.}~\bibnamefont
  {M{\"u}ller}}, \bibinfo {author} {\bibfnamefont {S.}~\bibnamefont
  {Bounouar}}, \bibinfo {author} {\bibfnamefont {K.~D.}\ \bibnamefont
  {J{\"o}ns}}, \bibinfo {author} {\bibfnamefont {M.}~\bibnamefont
  {Gl{\"a}ssl}}, \ and\ \bibinfo {author} {\bibfnamefont {P.}~\bibnamefont
  {Michler}},\ }\href {\doibase 10.1038/nphoton.2013.377} {\bibfield  {journal}
  {\bibinfo  {journal} {Nature Photonics}\ }\textbf {\bibinfo {volume} {8}},\
  \bibinfo {pages} {224} (\bibinfo {year} {2014})}\BibitemShut {NoStop}%
\bibitem [{\citenamefont {Winik}\ \emph {et~al.}(2017)\citenamefont {Winik},
  \citenamefont {Cogan}, \citenamefont {Don}, \citenamefont {Schwartz},
  \citenamefont {Gantz}, \citenamefont {Schmidgall}, \citenamefont {Livneh},
  \citenamefont {Rapaport}, \citenamefont {Buks},\ and\ \citenamefont
  {Gershoni}}]{Roni.Winik.2017}%
  \BibitemOpen
  \bibfield  {author} {\bibinfo {author} {\bibfnamefont {R.}~\bibnamefont
  {Winik}}, \bibinfo {author} {\bibfnamefont {D.}~\bibnamefont {Cogan}},
  \bibinfo {author} {\bibfnamefont {Y.}~\bibnamefont {Don}}, \bibinfo {author}
  {\bibfnamefont {I.}~\bibnamefont {Schwartz}}, \bibinfo {author}
  {\bibfnamefont {L.}~\bibnamefont {Gantz}}, \bibinfo {author} {\bibfnamefont
  {E.~R.}\ \bibnamefont {Schmidgall}}, \bibinfo {author} {\bibfnamefont
  {N.}~\bibnamefont {Livneh}}, \bibinfo {author} {\bibfnamefont
  {R.}~\bibnamefont {Rapaport}}, \bibinfo {author} {\bibfnamefont
  {E.}~\bibnamefont {Buks}}, \ and\ \bibinfo {author} {\bibfnamefont
  {D.}~\bibnamefont {Gershoni}},\ }\href {\doibase 10.1103/PhysRevB.95.235435}
  {\bibfield  {journal} {\bibinfo  {journal} {Phys. Rev. B}\ }\textbf {\bibinfo
  {volume} {95}},\ \bibinfo {pages} {235435} (\bibinfo {year}
  {2017})}\BibitemShut {NoStop}%
\bibitem [{\citenamefont {Liu}\ \emph {et~al.}(2019)\citenamefont {Liu},
  \citenamefont {Su}, \citenamefont {Wei}, \citenamefont {Yao}, \citenamefont
  {Silva}, \citenamefont {Yu}, \citenamefont {Iles-Smith}, \citenamefont
  {Srinivasan}, \citenamefont {Rastelli}, \citenamefont {Li},\ and\
  \citenamefont {Wang}}]{Liu2019}%
  \BibitemOpen
  \bibfield  {author} {\bibinfo {author} {\bibfnamefont {J.}~\bibnamefont
  {Liu}}, \bibinfo {author} {\bibfnamefont {R.}~\bibnamefont {Su}}, \bibinfo
  {author} {\bibfnamefont {Y.}~\bibnamefont {Wei}}, \bibinfo {author}
  {\bibfnamefont {B.}~\bibnamefont {Yao}}, \bibinfo {author} {\bibfnamefont
  {S.~F. C.~d.}\ \bibnamefont {Silva}}, \bibinfo {author} {\bibfnamefont
  {Y.}~\bibnamefont {Yu}}, \bibinfo {author} {\bibfnamefont {J.}~\bibnamefont
  {Iles-Smith}}, \bibinfo {author} {\bibfnamefont {K.}~\bibnamefont
  {Srinivasan}}, \bibinfo {author} {\bibfnamefont {A.}~\bibnamefont
  {Rastelli}}, \bibinfo {author} {\bibfnamefont {J.}~\bibnamefont {Li}}, \ and\
  \bibinfo {author} {\bibfnamefont {X.}~\bibnamefont {Wang}},\ }\href {\doibase
  10.1038/s41565-019-0435-9} {\bibfield  {journal} {\bibinfo  {journal} {Nature
  Nanotechnology}\ }\textbf {\bibinfo {volume} {14}},\ \bibinfo {pages} {586}
  (\bibinfo {year} {2019})}\BibitemShut {NoStop}%
\bibitem [{\citenamefont {Wang}\ \emph {et~al.}(2019)\citenamefont {Wang},
  \citenamefont {Hu}, \citenamefont {Chung}, \citenamefont {Qin}, \citenamefont
  {Yang}, \citenamefont {Li}, \citenamefont {Liu}, \citenamefont {Zhong},
  \citenamefont {He}, \citenamefont {Ding}, \citenamefont {Deng}, \citenamefont
  {Dai}, \citenamefont {Huo}, \citenamefont {H\"ofling}, \citenamefont {Lu},\
  and\ \citenamefont {Pan}}]{Wang2019}%
  \BibitemOpen
  \bibfield  {author} {\bibinfo {author} {\bibfnamefont {H.}~\bibnamefont
  {Wang}}, \bibinfo {author} {\bibfnamefont {H.}~\bibnamefont {Hu}}, \bibinfo
  {author} {\bibfnamefont {T.-H.}\ \bibnamefont {Chung}}, \bibinfo {author}
  {\bibfnamefont {J.}~\bibnamefont {Qin}}, \bibinfo {author} {\bibfnamefont
  {X.}~\bibnamefont {Yang}}, \bibinfo {author} {\bibfnamefont {J.-P.}\
  \bibnamefont {Li}}, \bibinfo {author} {\bibfnamefont {R.-Z.}\ \bibnamefont
  {Liu}}, \bibinfo {author} {\bibfnamefont {H.-S.}\ \bibnamefont {Zhong}},
  \bibinfo {author} {\bibfnamefont {Y.-M.}\ \bibnamefont {He}}, \bibinfo
  {author} {\bibfnamefont {X.}~\bibnamefont {Ding}}, \bibinfo {author}
  {\bibfnamefont {Y.-H.}\ \bibnamefont {Deng}}, \bibinfo {author}
  {\bibfnamefont {Q.}~\bibnamefont {Dai}}, \bibinfo {author} {\bibfnamefont
  {Y.-H.}\ \bibnamefont {Huo}}, \bibinfo {author} {\bibfnamefont
  {S.}~\bibnamefont {H\"ofling}}, \bibinfo {author} {\bibfnamefont {C.-Y.}\
  \bibnamefont {Lu}}, \ and\ \bibinfo {author} {\bibfnamefont {J.-W.}\
  \bibnamefont {Pan}},\ }\href {\doibase 10.1103/PhysRevLett.122.113602}
  {\bibfield  {journal} {\bibinfo  {journal} {Phys. Rev. Lett.}\ }\textbf
  {\bibinfo {volume} {122}},\ \bibinfo {pages} {113602} (\bibinfo {year}
  {2019})}\BibitemShut {NoStop}%
\bibitem [{\citenamefont {Bennett}\ \emph {et~al.}(2016)\citenamefont
  {Bennett}, \citenamefont {Lee}, \citenamefont {Ellis}, \citenamefont {Meany},
  \citenamefont {Murray}, \citenamefont {Floether}, \citenamefont {Griffths},
  \citenamefont {Farrer}, \citenamefont {Ritchie},\ and\ \citenamefont
  {Shields}}]{Bennette2016}%
  \BibitemOpen
  \bibfield  {author} {\bibinfo {author} {\bibfnamefont {A.~J.}\ \bibnamefont
  {Bennett}}, \bibinfo {author} {\bibfnamefont {J.~P.}\ \bibnamefont {Lee}},
  \bibinfo {author} {\bibfnamefont {D.~J.~P.}\ \bibnamefont {Ellis}}, \bibinfo
  {author} {\bibfnamefont {T.}~\bibnamefont {Meany}}, \bibinfo {author}
  {\bibfnamefont {E.}~\bibnamefont {Murray}}, \bibinfo {author} {\bibfnamefont
  {F.~F.}\ \bibnamefont {Floether}}, \bibinfo {author} {\bibfnamefont {J.~P.}\
  \bibnamefont {Griffths}}, \bibinfo {author} {\bibfnamefont {I.}~\bibnamefont
  {Farrer}}, \bibinfo {author} {\bibfnamefont {D.~A.}\ \bibnamefont {Ritchie}},
  \ and\ \bibinfo {author} {\bibfnamefont {A.~J.}\ \bibnamefont {Shields}},\
  }\href {https://advances.sciencemag.org/content/2/4/e1501256} {\bibfield
  {journal} {\bibinfo  {journal} {Science Advances}\ }\textbf {\bibinfo
  {volume} {2}} (\bibinfo {year} {2016})}\BibitemShut {NoStop}%
\bibitem [{\citenamefont {M\"uller}\ \emph {et~al.}(2017)\citenamefont
  {M\"uller}, \citenamefont {Vural}, \citenamefont {Schneider}, \citenamefont
  {Rastelli}, \citenamefont {Schmidt}, \citenamefont {H\"ofling},\ and\
  \citenamefont {Michler}}]{Muller2017}%
  \BibitemOpen
  \bibfield  {author} {\bibinfo {author} {\bibfnamefont {M.}~\bibnamefont
  {M\"uller}}, \bibinfo {author} {\bibfnamefont {H.}~\bibnamefont {Vural}},
  \bibinfo {author} {\bibfnamefont {C.}~\bibnamefont {Schneider}}, \bibinfo
  {author} {\bibfnamefont {A.}~\bibnamefont {Rastelli}}, \bibinfo {author}
  {\bibfnamefont {O.~G.}\ \bibnamefont {Schmidt}}, \bibinfo {author}
  {\bibfnamefont {S.}~\bibnamefont {H\"ofling}}, \ and\ \bibinfo {author}
  {\bibfnamefont {P.}~\bibnamefont {Michler}},\ }\href {\doibase
  10.1103/PhysRevLett.118.257402} {\bibfield  {journal} {\bibinfo  {journal}
  {Phys. Rev. Lett.}\ }\textbf {\bibinfo {volume} {118}},\ \bibinfo {pages}
  {257402} (\bibinfo {year} {2017})}\BibitemShut {NoStop}%
\bibitem [{\citenamefont {Olson}\ \emph {et~al.}(2017)\citenamefont {Olson},
  \citenamefont {Motes}, \citenamefont {Birchall}, \citenamefont {Studer},
  \citenamefont {LaBorde}, \citenamefont {Moulder}, \citenamefont {Rohde},\
  and\ \citenamefont {Dowling}}]{Olson2017}%
  \BibitemOpen
  \bibfield  {author} {\bibinfo {author} {\bibfnamefont {J.~P.}\ \bibnamefont
  {Olson}}, \bibinfo {author} {\bibfnamefont {K.~R.}\ \bibnamefont {Motes}},
  \bibinfo {author} {\bibfnamefont {P.~M.}\ \bibnamefont {Birchall}}, \bibinfo
  {author} {\bibfnamefont {N.~M.}\ \bibnamefont {Studer}}, \bibinfo {author}
  {\bibfnamefont {M.}~\bibnamefont {LaBorde}}, \bibinfo {author} {\bibfnamefont
  {T.}~\bibnamefont {Moulder}}, \bibinfo {author} {\bibfnamefont {P.~P.}\
  \bibnamefont {Rohde}}, \ and\ \bibinfo {author} {\bibfnamefont {J.~P.}\
  \bibnamefont {Dowling}},\ }\href {\doibase 10.1103/PhysRevA.96.013810}
  {\bibfield  {journal} {\bibinfo  {journal} {Phys. Rev. A}\ }\textbf {\bibinfo
  {volume} {96}},\ \bibinfo {pages} {013810} (\bibinfo {year}
  {2017})}\BibitemShut {NoStop}%
\bibitem [{\citenamefont {Su}\ \emph {et~al.}(2017)\citenamefont {Su},
  \citenamefont {Li}, \citenamefont {Rohde}, \citenamefont {Huang},
  \citenamefont {Wang}, \citenamefont {Li}, \citenamefont {Liu}, \citenamefont
  {Dowling}, \citenamefont {Lu},\ and\ \citenamefont {Pan}}]{Zuen2017}%
  \BibitemOpen
  \bibfield  {author} {\bibinfo {author} {\bibfnamefont {Z.-E.}\ \bibnamefont
  {Su}}, \bibinfo {author} {\bibfnamefont {Y.}~\bibnamefont {Li}}, \bibinfo
  {author} {\bibfnamefont {P.~P.}\ \bibnamefont {Rohde}}, \bibinfo {author}
  {\bibfnamefont {H.-L.}\ \bibnamefont {Huang}}, \bibinfo {author}
  {\bibfnamefont {X.-L.}\ \bibnamefont {Wang}}, \bibinfo {author}
  {\bibfnamefont {L.}~\bibnamefont {Li}}, \bibinfo {author} {\bibfnamefont
  {N.-L.}\ \bibnamefont {Liu}}, \bibinfo {author} {\bibfnamefont {J.~P.}\
  \bibnamefont {Dowling}}, \bibinfo {author} {\bibfnamefont {C.-Y.}\
  \bibnamefont {Lu}}, \ and\ \bibinfo {author} {\bibfnamefont {J.-W.}\
  \bibnamefont {Pan}},\ }\href {\doibase 10.1103/PhysRevLett.119.080502}
  {\bibfield  {journal} {\bibinfo  {journal} {Phys. Rev. Lett.}\ }\textbf
  {\bibinfo {volume} {119}},\ \bibinfo {pages} {080502} (\bibinfo {year}
  {2017})}\BibitemShut {NoStop}%
\bibitem [{\citenamefont {Resch}\ \emph {et~al.}(2007)\citenamefont {Resch},
  \citenamefont {Pregnell}, \citenamefont {Prevedel}, \citenamefont
  {Gilchrist}, \citenamefont {Pryde}, \citenamefont {O'Brien},\ and\
  \citenamefont {White}}]{Resch2007}%
  \BibitemOpen
  \bibfield  {author} {\bibinfo {author} {\bibfnamefont {K.~J.}\ \bibnamefont
  {Resch}}, \bibinfo {author} {\bibfnamefont {K.~L.}\ \bibnamefont {Pregnell}},
  \bibinfo {author} {\bibfnamefont {R.}~\bibnamefont {Prevedel}}, \bibinfo
  {author} {\bibfnamefont {A.}~\bibnamefont {Gilchrist}}, \bibinfo {author}
  {\bibfnamefont {G.~J.}\ \bibnamefont {Pryde}}, \bibinfo {author}
  {\bibfnamefont {J.~L.}\ \bibnamefont {O'Brien}}, \ and\ \bibinfo {author}
  {\bibfnamefont {A.~G.}\ \bibnamefont {White}},\ }\href {\doibase
  10.1103/PhysRevLett.98.223601} {\bibfield  {journal} {\bibinfo  {journal}
  {Phys. Rev. Lett.}\ }\textbf {\bibinfo {volume} {98}},\ \bibinfo {pages}
  {223601} (\bibinfo {year} {2007})}\BibitemShut {NoStop}%
\bibitem [{\citenamefont {Nagata}\ \emph {et~al.}(2007)\citenamefont {Nagata},
  \citenamefont {Okamoto}, \citenamefont {O{\textquoteright}Brien},
  \citenamefont {Sasaki},\ and\ \citenamefont {Takeuchi}}]{Nagata2007}%
  \BibitemOpen
  \bibfield  {author} {\bibinfo {author} {\bibfnamefont {T.}~\bibnamefont
  {Nagata}}, \bibinfo {author} {\bibfnamefont {R.}~\bibnamefont {Okamoto}},
  \bibinfo {author} {\bibfnamefont {J.~L.}\ \bibnamefont
  {O{\textquoteright}Brien}}, \bibinfo {author} {\bibfnamefont
  {K.}~\bibnamefont {Sasaki}}, \ and\ \bibinfo {author} {\bibfnamefont
  {S.}~\bibnamefont {Takeuchi}},\ }\href {\doibase 10.1126/science.1138007}
  {\bibfield  {journal} {\bibinfo  {journal} {Science}\ }\textbf {\bibinfo
  {volume} {316}},\ \bibinfo {pages} {726} (\bibinfo {year}
  {2007})}\BibitemShut {NoStop}%
\bibitem [{\citenamefont {Schwartz}\ \emph {et~al.}(2016)\citenamefont
  {Schwartz}, \citenamefont {Cogan}, \citenamefont {Schmidgall}, \citenamefont
  {Don}, \citenamefont {Gantz}, \citenamefont {Kenneth}, \citenamefont
  {Lindner},\ and\ \citenamefont {Gershoni}}]{cluster}%
  \BibitemOpen
  \bibfield  {author} {\bibinfo {author} {\bibfnamefont {I.}~\bibnamefont
  {Schwartz}}, \bibinfo {author} {\bibfnamefont {D.}~\bibnamefont {Cogan}},
  \bibinfo {author} {\bibfnamefont {E.~R.}\ \bibnamefont {Schmidgall}},
  \bibinfo {author} {\bibfnamefont {Y.}~\bibnamefont {Don}}, \bibinfo {author}
  {\bibfnamefont {L.}~\bibnamefont {Gantz}}, \bibinfo {author} {\bibfnamefont
  {O.}~\bibnamefont {Kenneth}}, \bibinfo {author} {\bibfnamefont {N.~H.}\
  \bibnamefont {Lindner}}, \ and\ \bibinfo {author} {\bibfnamefont
  {D.}~\bibnamefont {Gershoni}},\ }\href {\doibase 10.1126/science.aah4758}
  {\bibfield  {journal} {\bibinfo  {journal} {Science}\ }\textbf {\bibinfo
  {volume} {354}},\ \bibinfo {pages} {434} (\bibinfo {year}
  {2016})}\BibitemShut {NoStop}%
\bibitem [{\citenamefont {Schwartz}\ \emph {et~al.}(2015)\citenamefont
  {Schwartz}, \citenamefont {Schmidgall}, \citenamefont {Gantz}, \citenamefont
  {Cogan}, \citenamefont {Bordo}, \citenamefont {Don}, \citenamefont
  {Zielinski},\ and\ \citenamefont {Gershoni}}]{dark}%
  \BibitemOpen
  \bibfield  {author} {\bibinfo {author} {\bibfnamefont {I.}~\bibnamefont
  {Schwartz}}, \bibinfo {author} {\bibfnamefont {E.~R.}\ \bibnamefont
  {Schmidgall}}, \bibinfo {author} {\bibfnamefont {L.}~\bibnamefont {Gantz}},
  \bibinfo {author} {\bibfnamefont {D.}~\bibnamefont {Cogan}}, \bibinfo
  {author} {\bibfnamefont {E.}~\bibnamefont {Bordo}}, \bibinfo {author}
  {\bibfnamefont {Y.}~\bibnamefont {Don}}, \bibinfo {author} {\bibfnamefont
  {M.}~\bibnamefont {Zielinski}}, \ and\ \bibinfo {author} {\bibfnamefont
  {D.}~\bibnamefont {Gershoni}},\ }\href {\doibase 10.1103/PhysRevX.5.011009}
  {\bibfield  {journal} {\bibinfo  {journal} {Phys. Rev. X}\ }\textbf {\bibinfo
  {volume} {5}},\ \bibinfo {pages} {011009} (\bibinfo {year}
  {2015})}\BibitemShut {NoStop}%
\bibitem [{\citenamefont {Cogan}\ \emph {et~al.}(2018)\citenamefont {Cogan},
  \citenamefont {Kenneth}, \citenamefont {Lindner}, \citenamefont {Peniakov},
  \citenamefont {Hopfmann}, \citenamefont {Dalacu}, \citenamefont {Poole},
  \citenamefont {Hawrylak},\ and\ \citenamefont
  {Gershoni}}]{Cogan.Depolarization}%
  \BibitemOpen
  \bibfield  {author} {\bibinfo {author} {\bibfnamefont {D.}~\bibnamefont
  {Cogan}}, \bibinfo {author} {\bibfnamefont {O.}~\bibnamefont {Kenneth}},
  \bibinfo {author} {\bibfnamefont {N.~H.}\ \bibnamefont {Lindner}}, \bibinfo
  {author} {\bibfnamefont {G.}~\bibnamefont {Peniakov}}, \bibinfo {author}
  {\bibfnamefont {C.}~\bibnamefont {Hopfmann}}, \bibinfo {author}
  {\bibfnamefont {D.}~\bibnamefont {Dalacu}}, \bibinfo {author} {\bibfnamefont
  {P.~J.}\ \bibnamefont {Poole}}, \bibinfo {author} {\bibfnamefont
  {P.}~\bibnamefont {Hawrylak}}, \ and\ \bibinfo {author} {\bibfnamefont
  {D.}~\bibnamefont {Gershoni}},\ }\href {\doibase 10.1103/PhysRevX.8.041050}
  {\bibfield  {journal} {\bibinfo  {journal} {Phys. Rev. X}\ }\textbf {\bibinfo
  {volume} {8}},\ \bibinfo {pages} {041050} (\bibinfo {year}
  {2018})}\BibitemShut {NoStop}%
\bibitem [{\citenamefont {Schmidgall}\ \emph {et~al.}(2015)\citenamefont
  {Schmidgall}, \citenamefont {Schwartz}, \citenamefont {Cogan}, \citenamefont
  {Gantz}, \citenamefont {Heindel}, \citenamefont {Reitzenstein},\ and\
  \citenamefont {Gershoni}}]{Emma.APL.2015}%
  \BibitemOpen
  \bibfield  {author} {\bibinfo {author} {\bibfnamefont {E.~R.}\ \bibnamefont
  {Schmidgall}}, \bibinfo {author} {\bibfnamefont {I.}~\bibnamefont
  {Schwartz}}, \bibinfo {author} {\bibfnamefont {D.}~\bibnamefont {Cogan}},
  \bibinfo {author} {\bibfnamefont {L.}~\bibnamefont {Gantz}}, \bibinfo
  {author} {\bibfnamefont {T.}~\bibnamefont {Heindel}}, \bibinfo {author}
  {\bibfnamefont {S.}~\bibnamefont {Reitzenstein}}, \ and\ \bibinfo {author}
  {\bibfnamefont {D.}~\bibnamefont {Gershoni}},\ }\href {\doibase
  10.1063/1.4921000} {\bibfield  {journal} {\bibinfo  {journal} {Applied
  Physics Letters}\ }\textbf {\bibinfo {volume} {106}},\ \bibinfo {pages}
  {193101} (\bibinfo {year} {2015})}\BibitemShut {NoStop}%
\bibitem [{\citenamefont {Cogan}\ \emph {et~al.}(2020)\citenamefont {Cogan},
  \citenamefont {Peniakov}, \citenamefont {Su},\ and\ \citenamefont
  {Gershoni}}]{StateTomography.2020}%
  \BibitemOpen
  \bibfield  {author} {\bibinfo {author} {\bibfnamefont {D.}~\bibnamefont
  {Cogan}}, \bibinfo {author} {\bibfnamefont {G.}~\bibnamefont {Peniakov}},
  \bibinfo {author} {\bibfnamefont {Z.-E.}\ \bibnamefont {Su}}, \ and\ \bibinfo
  {author} {\bibfnamefont {D.}~\bibnamefont {Gershoni}},\ }\href {\doibase
  10.1103/PhysRevB.101.035424} {\bibfield  {journal} {\bibinfo  {journal}
  {Phys. Rev. B}\ }\textbf {\bibinfo {volume} {101}},\ \bibinfo {pages}
  {035424} (\bibinfo {year} {2020})}\BibitemShut {NoStop}%
\end{thebibliography}

\end{document}